\documentclass[thmsa,12pt]{article}
\normalfont\sffamily
\usepackage{amsfonts}
\usepackage{amssymb}
\usepackage[active]{srcltx}
\usepackage{color}
\usepackage[x11names]{xcolor}
\usepackage{amsmath}
\usepackage{mathtools}
\usepackage{amsthm}
\usepackage{bm} % for bold Greek symbols
\usepackage{graphicx}
\usepackage{textcomp}
\usepackage{a4wide}
\usepackage{chicago}
\usepackage{multirow}
\usepackage{diagbox}
\usepackage{color}
\usepackage{makecell}
\usepackage{longtable}
\usepackage{cellspace}
\usepackage{tabularx}
\usepackage{accents}
\usepackage{array,booktabs,ragged2e}
\usepackage{pxfonts}
\usepackage[format=hang]{caption}
\usepackage{subcaption}
\usepackage{url}
\usepackage{xfrac}

\usepackage[normalem]{ulem}
\usepackage{comment}
\usepackage{pxfonts}
\usepackage[margin=3cm]{geometry} % margin wird weiter unten wieder überschrieben
\usepackage{titlesec}

\usepackage{setspace}
\usepackage{footmisc}
\usepackage{accents}
\usepackage{appendix}

\usepackage{array,booktabs,ragged2e}
\newcolumntype{R}[1]{>{\RaggedRight}p{#1}}

\usepackage{cellspace}
\usepackage{hyperref}
\def \be {\begin{equation}}
\def \ee {\end{equation}}

\DeclareMathOperator*{\argmax}{arg\,max}

\begin{document}

\def\title #1{\begin{center}
{\Large {\sc #1}}
\end{center}}
\def\author #1{\begin{center} {#1}
\end{center}}

\setstretch{1.1}

\begin{titlepage}
    \phantomsection \label{Titlepage}
    \addcontentsline{toc}{section}{Title page}

\renewcommand{\thefootnote}{\fnsymbol{footnote}}\addtocounter{footnote}{1}
\title{\sc 
Influence in Weighted Committees 
\\ \medskip \  }

\author{Sascha Kurz\\ {\small Dept.\ of Mathematics, University of Bayreuth, Germany\\sascha.kurz@uni-bayreuth.de}}

\author{Alexander Mayer\\ {\small Dept.\ of Economics, University of Bayreuth, Germany\\alexander.mayer@uni-bayreuth.de  }}

\author{Stefan Napel\\ {\small Dept.\ of Economics, University of Bayreuth, Germany\\stefan.napel@uni-bayreuth.de }}

\vspace{0.3cm}

%% \begin{center} {\tt This draft: 
%% \hspace{-1.9em} \today} \end{center}

\vspace{0.2cm}
%\vspace{0.3cm}
%\vspace{4.01cm}

\begin{center} {\bf {\sc Abstract}} \end{center}
{\small 
A committee's decisions on %$m\ge 3$ 
more than two alternatives much depend on the adopted voting method, and so does the distribution of power among the committee members. 
We investigate how different aggregation methods such as plurality runoff, Borda count, or Copeland rule map asymmetric numbers of seats, shares, voting weights, etc.\ to influence on outcomes when preferences vary. 
A generalization of the Penrose-Banzhaf power index is proposed and applied to the IMF Executive Board's election of a Managing Director, extending a~priori voting power analysis from binary simple voting games to choice in weighted committees.
} 
\vspace{0.2cm}

\begin{description}
{\small
\item[Keywords:]
weighted voting $\cdot$ voting power $\cdot$ weighted committee games $\cdot$ plurality runoff $\cdot$ 
Borda rule~$\cdot$ Copeland rule  $\cdot$ Schulze rule $\cdot$
 IMF Executive Board $\cdot$ IMF Director
%\item[JEL codes:] D71 $\cdot$ C71 $\cdot$ C63 
% C63: Computational Techniques, Simulation Modeling
% C71: Cooperative Games
% D71: Social Choice, Clubs, Committees, Associations
}
\end{description}

\vspace{1.6cm}

\vfill

\noindent {\footnotesize We are grateful to Hannu Nurmi for stimulating discussions on the topic. We also benefitted from feedback on seminar or workshop presentations in Bamberg, Bayreuth, Berlin, Bremen, Dagstuhl, Delmenhorst, Graz, Hagen, Hamburg, Hanover, Leipzig, Moscow, Munich and Turku.}

\end{titlepage}

\addtocounter{footnote}{-1}
%\newpage
%\pagenumbering{arabic}
%\addtocounter{footnote}{-2}

%%%%%%%%%%%%%%%%%%%%%%%%%%%%%%%%%%%%%%%%%%%%%%%%%%%%%%%%%%%%%%%%%%%%%%%%%
\setstretch{1.2}

\pagenumbering{arabic}

%%%%%%%%%%%%%%%%%%%%
%%%%%%%%%%%%%%%%%%%%
%%%%%%%%%%%%%%%%%%%% INTRODUCTION
%%%%%%%%%%%%%%%%%%%%
%%%%%%%%%%%%%%%%%%%%

\section{Introduction}\label{sec:Introduction}

The aggregation of individual preferences by some form of voting is common in politics, business, and everyday life. Members of a board, council, or committee are rarely aware how much their collective choices depend on the adopted aggregation rule.
The importance of the method may be identified a~posteriori by comparing voting outcomes for a given preference configuration. 
For instance, suppose a hiring committee involved three groups with $n_1=6$, $n_2=5$, and $n_3=3$ members each and the following preferences over five candidates $\{a,b,c,d,e\}$: $a \succ_1 d \succ_1 e \succ_1 c \succ_1 b $, % for group~1, 
$b \succ_2 c \succ_2 d \succ_2 e \succ_2 a$, %for group~2, 
and $c \succ_3 e \succ_3 d \succ_3 b \succ_3 a$. % for group 3. 
If the groups voted sincerely (for informational, institutional, or other reasons) then candidate $a$ would have received the position under \emph{plurality rule} with 6 vs.\ 5 vs.\ 3 votes. However, %the popular 
%\emph{plurality voting with a runoff} 
requiring a \emph{runoff} between the front-runners, given that none has majority support, would have made $b$ the winner with 8 to 6 votes. Candidate $c$ would have won every pairwise comparison and been the \emph{Condorcet winner}; \emph{Borda rule} would have singled out $d$; candidate $e$ could have been the winner if \emph{approval voting} had been used.  

With enough posterior information, each voter group can identify a `most-prefer\-red voting method' for the decision at hand: group~1 should have tried to impose plurality rule in order to have its way;  group~2 should have argued for plurality runoff; and group~3 for pairwise comparisons. 
It is less obvious, though, how adoption of one aggregation method rather than another will affect a group's success and influence \emph{a~priori}, i.e., not yet knowing what will be the applicable preferences, or %if one averages over 
averaged across many %similar 
decisions. 
More generally, can we verify if small groups are enjoying a greater say when committees fill a top position, % or 
elect an official, etc.\ by pairwise votes or by the plurality runoff method? Which rules from a given list of suggestions tend to maximize (or minimize) voting power of a particularly sized group in a committee?
There is a huge literature on voting power but questions of this kind have to our knowledge not been addressed by it yet.\footnote{The most closely related analysis seems to be the investigation of effectivity functions; cf.\ \citeN{Peleg:1984}. See \citeN{Felsenthal/Machover:1998}, \citeN{Laruelle/Valenciano:2008}, or \citeN{Napel:2018} for overviews on the measurement of voting power.}
We seek to change this by generalizing tools that were developed for analysis of {simple voting games} with binary options (`yes' or `no') to collective choice from $m\ge 3$ alternatives. 

The most prominent tools for analyzing the former are the \emph{Penrose-Banzhaf index} (\citeNP{Penrose:1946}; \citeNP{Banzhaf:1965}) and the \emph{Shapley-Shubik index} \cite{Shapley/Shubik:1954}. They evaluate the sensitivity of the outcome to changes in a given voter's preferences -- operationalized as the likelihood of this voter being \emph{pivotal} or \emph{critical}: flipping its vote would swing the collective decision -- under specific probabilistic assumptions. This paper applies the same logic to {weighted committee games}.
These, more simply referred to as \emph{weighted committees}, are %defined in Kurz, Mayer, and Napel~(\citeyear{Kurz/Mayer/Napel:2018}) as 
tuples $(N,A, r|\mathbf{w})$ that specify a set $N=\{1, \ldots, n\}$ of players, a set $A=\{a_1, \ldots, a_m\}$ of alternatives, and the combination $r|\mathbf{w}$ of an anonymous voting method $r$ (e.g., Borda rule, plurality rule, and so on) with a vector $\mathbf{w}$ of integers that represents group sizes, voting shares, etc. 
%Formally, $r$ is a family of mappings from profiles $\mathbf{P}=(P_1, \ldots, P_n)$ of individual preferences over $A$ to $A$ that is invariant under permutations %$\pi\colon 
%$N\to N$; and voting weights $w_1, \ldots, w_n$ replicate each preference $P_i$ in  $r$'s argument $w_i$ times:
%$r|\mathbf{w}(\mathbf{P})=r([P_1]^{w_1}, [P_2]^{w_2}, \ldots, [P_n]^{w_n})$ with $[P_i]^{w_i}:=(P_i, \ldots, P_i)$. 

As in analysis of (binary) simple voting games, high \emph{a priori voting power} of player~$i\in N$ in a committee goes with high sensitivity of the outcome to $i$'s preferences. 
It can be quantified as the probability for a change in $i$'s preferences causing a change of the collective choice. 
Although other assumptions could be made, we focus on the simplest case in which all profiles of strict preference orderings are equally likely a~priori.
This corresponds  to the Penrose-Banzhaf index for $m=2$. 
The respective power indications %can 
help to assess who gains and loses from institutional reforms or whether the distribution of influence in a committee satisfies some fairness criterion; %desiderata;  
they can also inform stakeholders, lobbyists, or other committee outsiders with an interest in who has how much say in the committee. 

%We briefly review related work and introduce notation %and a few voting rules of interest in
%%(Sections~\ref{sec:literature} and \ref{sec:preliminaries}) 
%before we %get to 
%quantify %define our new measure of 
%influence in weighted committees. %(Section~\ref{sec:influence_measure}). 
%The proposed power measure will be illustrated by a toy example and %by evaluating
%2016's voting weight reform for the Executive Board of the International Monetary Fund. %(Section~\ref{sec:illustration}).

\section{Related Work}
The distribution of power in binary weighted voting games has received wide attention ever since \citeN[Ch.~10]{vonNeumann/Morgenstern:1953} formalized them as a subclass of so-called \emph{simple (voting) games}. See, e.g., \citeN{Mann/Shapley:1962}, \citeN{Riker/Shapley:1968}, \citeN{Owen:1975:presidential}, or \citeN{Brams:1978} for seminal investigations.  %\citeN{Felsenthal/Machover:1998}, \citeN{Laruelle/Valenciano:2008}, or \citeN{Napel:2018} review the definitions and different interpretations of  Shapley-Shubik index \cite{Shapley/Shubik:1954}, Penrose-Banzhaf index (\citeNP{Penrose:1946}, \citeNP{Banzhaf:1965}), and other power indices for simple voting games.
The binary framework can be restrictive, however.
Even for collective `yes'-or-`no' decisions, individual voters usually have more than two options. For instance they can abstain or not even attend a vote, and this may affect the outcome differently than casting a vote either way. 
Corresponding situations have been formalized as \emph{ternary voting games} (\citeNP{Felsenthal/Machover:1997}; \citeNP{Tchantcho/DiffoLambo/Pongou/Engoulou:2008}; \citeNP{Parker:2012}) and \emph{quaternary voting games} \cite{Laruelle/Valenciano:2012}.
Players can also be allowed to express graded intensities of support: in \emph{$(j,k)$-games}, studied by \citeN{Hsiao/Raghavan:1993} and Freixas and Zwicker~(\citeyearNP{Freixas/Zwicker:2003}, \citeyearNP{Freixas/Zwicker:2009}), each player selects one of $j$ ordered levels of approval. The resulting $j$-partitions of players are mapped to one of $k$ ordered output levels; 
suitable power indices %for simple voting games (that is, $(2,2)$-games) have been extended to general $(j,k)$-games 
have been defined by Freixas~(\citeyearNP{Freixas:2005:Banzhaf}, \citeyearNP{Freixas:2005:Shapley}).  

Linear orderings of actions and feasible outcomes, as required by $(j,k)$-games, are naturally given in many applications but fail to exist in others. 
Think of options that have multidimensional attributes -- for instance, candidates for office or an open position, policy programs, locations of a facility, etc.
Pertinent extensions of simple games, along with corresponding power measures, have been introduced as \emph{multicandidate voting games} %in a series of papers by Bolger (see, e.g., \citeyearNP{Bolger:1980,Bolger:1986,Bolger:2000,Bolger:2002}) 
by \citeN{Bolger:1986}
and taken up as \emph{simple $r$-games} % by Amer et al.\ (\citeyearNP{Amer/Carreras/Magana:1998a,Amer/Carreras/Magana:1998b}) 
by \citeN{Amer/Carreras/Magana:1998b}
and as \emph{weighted plurality games} by \citeN{Chua/Ueng/Huang:2002}. 
They require each player to vote for a single candidate. This results in partitions of player set $N$ that, in contrast to $(j,k)$-games, are mapped to a winning candidate without ordering restrictions. 

We will draw on the yet more general framework of \emph{weighted committee games} (\citeNP{Kurz/Mayer/Napel:2018}). Winners in these games can depend on the entire preference rankings of voters rather than just the respective top.
We conceive of \emph{player~$i$'s influence} %in the respective decision-making body 
or \emph{voting power} as the sensitivity of joint decisions to $i$'s actions or likings. 
The resulting ability to affect collective outcomes is closely linked to the opportunity to manipulate social choices in the sense of \citeN{Gibbard:1973} and \citeN{Satterthwaite:1975}. 
Our investigation therefore relates to computational studies by \citeN{Nitzan:1985}, \citeN{Kelly:1993}, \citeN{Aleskerov/Kurbanov:1999}, or \citeN{Smith:1999} that have quantified the aggregate \emph{manipulability} of a given decision rule. 
The %key 
conceptual difference between corresponding manipulability indices and the power index defined below is that the latter is evaluating consequences of arbitrary preference perturbations, while the indicated studies only look at strategic preference misrepresentation that is beneficial from the perspective of a player's original preferences.\footnote{\citeN{Nitzan:1985} also checked if outcomes could be affected by arbitrary variations of %individual
preferences before assessing manipulation. He tracked this at the aggregate level, while we break it down to individuals in order to %assess their power and 
link outcome sensitivity to voting weights.} 
Voting power as we quantify it could be used for a %to the considered 
player's strategic advantage but it need not. 
A `preference change' might also be purely idiosyncratic, result from log-rolling or external lobbying (where costs of persuasion 
% or the price of buying votes 
can relate more to preference intensity than a player's original ranking of options), or could be a demonstration of power for its own sake. 

Other conceptualizations of the influence derived from a given collective choice rule track the sets of outcomes that can be induced by partial coalitions. 
For instance, \citeN{Moulin:1981} uses \emph{veto functions} %for neutral rules in order 
in order to describe outcomes that given coalitions of players could jointly prevent; % if they coordinated. % their behavior. 
Peleg's \citeyear{Peleg:1984} \emph{effectivity functions} describe the power structure in a committee by a list of all sets of alternatives that specific coalitions of %(either sincere or strategic) 
voters can force the outcome to lie in. % (assuming either no or perfect information about others). 
We, by contrast, follow the literature pioneered by \citeN{Penrose:1946}, \citeN{Shapley/Shubik:1954}, and \citeN{Banzhaf:1965}, and try to assess individual influence on outcomes concisely by a number between zero and one.

\section{Preliminaries\label{sec:preliminaries}}
%\subsection{Notation and definitions\label{sec:notation}}
\subsection{Anonymous Voting Rules\label{sec:rules}}
We consider finite sets $N=\{1,\dots,n\}$ of $n$ voters or players such that each voter~$i\in N$ has strict preferences $P_i$ over a set $A=\{a_1,\dots,a_m\}$ of $m\ge 2$ alternatives. We write $abc$ in abbreviation of $aP_ibP_ic$ when the player's identity is clear. The set of all $m!$ strict preference orderings on $A$ is denoted by $\mathcal{P}(A)$. A \emph{(resolute) voting rule} $r\colon  \mathcal{P}(A)^n \to A$ maps each preference profile $\mathbf{P}=(P_1,\dots,P_n)$ to a single winning alternative $a^*=r(\mathbf{P})$. Rule $r$  is \emph{anonymous} if for any $\mathbf{P}\in \mathcal{P}(A)$ and any permutation $\pi\colon N\to N$ with $\pi(\mathbf{P})=(P_{\pi(1)},  \dots, P_{\pi(n)})$ we have $r(\mathbf{P})=r(\pi(\mathbf{P}))$.

\renewcommand{\arraystretch}{1.4}
\begin{table}
	\begin{center}
		\begin{tabular}{|l|l|}
			\hline\hline
			\emph{Rule} &\emph{ Winning alternative at preference profile $\mathbf{P}$ }\\
			\hline  \hline
			%			Anti-plurality %$r^A$ 
			%			&  $r^A(\mathbf{P})\in \argmin_{a\in A} \big|\{i\in N \ |\ \forall a'\neq a\in A\colon a' P_i a\}\big|$
			%			\\
			Borda %$r^B$ 
			&  $r^B(\mathbf{P})\in \argmax_{a\in A} \sum_{i\in N} b_i(a,\mathbf{P})$\\ \hline
			Copeland %$r^C$ 
			& $r^C(\mathbf{P})\in \argmax_{a\in A} \big|\{a'\in A \ |\ a \succ_M^\mathbf{P} a'\}\big|$ \\ \hline
			Plurality %$r^P$ 
			&  $r^P(\mathbf{P})\in \argmax_{a\in A} \big|\{i\in N \ |\ \forall a'\neq a\in A\colon a P_i a'\}\big|$ \\
			 \hline
			 & \\[-0.6cm]
			Plurality runoff 
			&  
				$r^{PR}(\mathbf{P}) 
			\begin{dcases}
			= r^P(\mathbf{P}) \; \text{ if } \; \big|\{i\in N \ |\ \forall a'\in A\setminus\{ r^P(\mathbf{P})\}\colon r^P(\mathbf{P}) P_i a'\}\big| > \tfrac{n}{2}\\
			\in \argmax_{a\in \{a_{(1)}, a_{(2)}\}} \big|\{i\in N \ |\ \forall a'\neq a\in \{a_{(1)}, a_{(2)}\} \colon a P_i a'\}\big|  \text{ otherwise} 
			\end{dcases}$ 
			\\[0.55cm] \hline
			Schulze & $r^S(\mathbf{P})$ -- see \citeN{Schulze:2011} \\
			\hline \hline
		\end{tabular} 
		\caption{Considered anonymous voting rules \label{table:rules}}
	\end{center}
\end{table}

We will restrict attention to truthful voting\footnote{In principle, power analysis could also be carried out for strategic voters. This would require specifying the mapping from profiles of players' preferences to the element of $A$ (or a probability distribution over $A$) which is induced by the selected voting equilibrium. Determination of the latter usually is a hard task in itself and here left aside.} under one of the five anonymous rules summarized in Table~\ref{table:rules}, assuming {lexicographic tie breaking}. 
See %, e.g., %\citeN{Nurmi:2006} or 
\citeN{Laslier:2012} on the pros and cons of a big variety of voting procedures. Our selection comprises two positional rules (Borda, plurality), two Condorcet methods (Copeland, Schulze), and a two-stage procedure that is used for filling political offices in many European jurisdictions (plurality runoff). 

Under \emph{plurality rule} $r^P$, each voter simply names his or her top-ranked alternative and the alternative that is ranked first by the most voters is chosen.
This is also the winner under \emph{plurality with runoff rule}~$r^{PR}$ if the obtained plurality constitutes a majority (i.e., more than 50\% of votes); otherwise a binary runoff between the alternatives $a_{(1)}$ and $a_{(2)}$ that obtained the highest and second-highest plurality scores in the first stage is conducted.

\emph{Borda rule} $r^B$ has each player~$i$ assign $m-1$, $m-2$, \ldots, $0$ points to the alternative that he or she ranks first, second, etc. These points
$
b_i(a,\mathbf{P}):= \big|\{a'\in A \ |\ a P_i a'\}\big|
$
equal the number of alternatives that $i$ ranks below $a$. The alternative with the highest total number of points, known as its {Borda score}, is selected. %Finally, 

%While the first three rules belong to the class of scoring rules, 
\emph{Copeland rule} $r^C$ considers pairwise majority votes between the alternatives. They % latter 
define the {majority relation} %$\succ_M^\mathbf{P}$ by 
$
a\succ_M^\mathbf{P} a' \ :\Leftrightarrow \  \big|\{i\in N\ | \ a P_i a' \} \big| >  \big|\{i\in N\ | \ a' P_i a \}\big|
$
and the alternative that beats the most others according to $\succ_M^\mathbf{P}$ is selected. 
Copeland rule is a {Condorcet method}: if some alternative $a$ % is a Condorcet winner, i.e., 
beats all others, 
then $r^C(\mathbf{P})=a$. 
The same is true for \emph{Schulze rule} $r^S$.  But 
%Let it suffice to note that, w
while $r^C$ just counts the number of direct pairwise victories, $r^S$ also considers indirect victories and invokes majority margins in order to evaluate their strengths. $r^S$ then picks the alternative that %, informally speaking, 
has the strongest chain of direct or indirect majority support  %, even though it might lose some direct comparisons (similar to Simpson-Kramer maximin rule, but avoiding some of the latter's flaws) and 
(%its computation is more elaborate and we refer to 
%see \citet{Schulze:2011} for details). 
see \citeNP{Schulze:2011} for details).
The attention paid to margins makes $r^S$ more sensitive to %individual 
voting weights than $r^C$ in case $\succ_M^\mathbf{P}$ is cyclical.  
Despite its non-trivial computation, $r^S$ has been applied, e.g., by the Wikimedia Foundation and Linux open-source communities. %, and Pirate Parties in Europe.

\subsection{Weighted Committees\label{sec:wcg}}

Anonymous rules treat any components $P_i$ and $P_j$ of a preference profile $\mathbf{P}$ like indistinguishable ballots. 
Still, when a committee conducts pairwise comparisons, plurality voting, etc., %a voting method like any of the above, 
two individual preferences often %can still 
feed into the collective decision rather asymmetrically because, e.g., stockholders have as many votes as they own shares or the relevant players $i\in N$ in a political committee %are parties that 
cast bloc votes in proportion to party seats. The resulting (non-anonymous) mapping from preferences to outcomes is a combination of anonymous voting rule $r$ with weights $w_1, \ldots, w_n$ for players $1, \ldots,n$ that is defined for all  $\mathbf{P}\in \mathcal{P}(A)^n$ by
\begin{equation}
r|\mathbf{w}(\mathbf{P}):=r([P_1]^{w_1}, [P_2]^{w_2}, \ldots, [P_n]^{w_n})=r(\underbrace{P_1, \ldots, P_1}_{w_1 \text{ times}}, \underbrace{P_2, \ldots, P_2}_{w_2 \text{ times}},\ldots, \underbrace{P_n, \ldots, P_n}_{w_n \text{ times}}). 
%\text{ for all } \mathbf{P}\in \mathcal{P}(A)^n.
\end{equation}
%for every profile of preferences over $A=\{a_1, \ldots, a_m\}$. 
The combination $(N,A,r|\mathbf{w})$ of a set of voters, a set of alternatives and a particular weighted voting rule defines a \emph{weighted committee (game)}. When the underlying anonymous rule is plurality rule $r^P$, then $(N,A,r^P|\mathbf{w})$ is called a \emph{(weighted) plurality committee}. Similarly, $(N,A,r^{PR}|\mathbf{w})$,  $(N,A,r^B|\mathbf{w})$, $(N,A,r^C|\mathbf{w})$ and $(N,A,r^S|\mathbf{w})$ are referred to as a \emph{plurality runoff committee}, \emph{Borda committee}, \emph{Copeland committee}, and \emph{Schulze committee}. 
See \shortciteN{Kurz/Mayer/Napel:2018} on structural differences between them. %properties of the rules and, e.g., stark differences in the numbers of distinct plurality and Borda committees. % for given $m, n\ge 3$.

Weighted committee games  $(N, A,r|\mathbf{w})$ and $(N, A,r'|\mathbf{w'})$ are \emph{equivalent} if the respective mappings from preference profiles %$\mathbf{P}$ 
to outcomes $a^*$ coincide; that is, when $r|\mathbf{w}(\mathbf{P}) = r'|\mathbf{w'}(\mathbf{P})$ for all $\mathbf{P}\in \mathcal{P}(A)^n$. 
Equivalent games evidently come with equivalent expectations for individual players to influence the collective decision (voting power) and to obtain outcomes that match their own preferences (success). 
We will focus on voting power and non-equivalent committees that involve either the same rule $r$ but different weights $\mathbf{w}$ and $\mathbf{w'}$, or the same weights $\mathbf{w}$ but different rules $r$ and $r'$. 
Example questions that we would like to address are:
to what extent does a change of voting weights, implied for example by the recent re-allocation of voting rights in the International Monetary Fund or party-switching of a member of parliament, shift the respective balance of power? 
How does players' attractiveness to lobbyists change when a committee replaces one voting method by another?

\section{Measuring Influence in Weighted Committees\label{sec:influence_measure}}
%\subsection{General Idea\label{sec:method}}

Some obvious shortcomings notwithstanding (see, e.g., \citeNP{Garrett/Tsebelis:1999}, \citeyearNP{Garrett/Tsebelis:2001}), voting power indices such as the Penrose-Banzhaf and the Shapley-Shubik index have received wide attention in theoretical and applied analysis of binary decisions. See, e.g., the contributions in \citeN{Holler/Nurmi:2013}. 
Most indices can be seen as operationalizing power of player~$i$ as \emph{expected sensitivity of collective decisions to player~$i$'s behavior} \cite{Napel/Widgren:2004}. 
Sensitivity in the binary case means that, taking the behavior of other players as given, the collective outcome would have been different had player~$i$ voted `no' instead of `yes', or `yes' instead of `no'. 
Distinct indices reflect distinct probabilistic assumptions about the voting configurations that are evaluated. 
For instance, the Penrose-Banzhaf index is predicated on the assumption that the %`yes'-or-`no' 
preferences of all $n$ players are independent random variables 
%with equal probabilities % $\sfrac{1}{2}$ 
%for `yes' and `no', i.e., 
and the $2^n$ different `yes'-`no' configurations %of $n$ players are all 
are equally likely. % in assessing sensitivity of the outcome to individual players $i\in N$. 

\subsection{Power as (Normalized) Expected Sensitivity of the Outcome}

It is not hard to generalize the idea of measuring influence as outcome sensitivity to weighted committees $(N, A,r|\mathbf{w})$. 
%under rule $r$ as the expected sensitivity of $r|\mathbf{w}(\mathbf{P})$ to changes of $P_i$. 
Continuing in the Penrose-Banzhaf tradition, we will assume that individual preferences are drawn independently from the uniform distribution over $\mathcal{P}(A)$, i.e., all profiles $\mathbf{P}\in \mathcal{P}(\{a_1, \ldots, a_m\})^n$ are equally likely.\footnote{This is also known as the \emph{impartial culture assumption}. It has limited empirical support (see, e.g., \citeNP[Ch.~1]{Regenwetter/Grofman/Marley/Tsetlin:2012}) but is commonly adopted as a starting point for assessing links between voting weights and power a~priori. We leave the pursuit of other options for future research (e.g., single-peakedness along a common dimension).}  
In order to assess the voting power of player~$i$ with weight $w_i$, we perturb $i$'s realized preferences $P_i$ to a random $P_i'\neq P_i \in \mathcal{P}(A)$ and check if this individual preference change would change the collective outcome.\footnote{One might also restrict attention to local %preference 
perturbations, that is, only allow changes of $P_i$ to adjacent orderings. So when $m=3$ and $P_i=abc$, one would impose the constraint $P_i'\in\{acb,bac\}$. This would not change the qualitative observations in the IMF section below.}
%Section~\ref{sec:IMF}.} 
Specifically, using notation  $\mathbf{P}=(P_i,\mathbf{P}_{-i})$ with $\mathbf{P}_{-i}= (P_1,\dots,P_{i-1},P_{i+1},\dots,P_n)$, we are interested in the behavior of indicator function
\begin{equation}
\Delta r|\mathbf{w}(\mathbf{P};{P'_i}):=
\begin{cases}
1& \text{if} \quad r|\mathbf{w}(\mathbf{P}) \neq r|\mathbf{w}(P_i',\mathbf{P}_{-i}), \\
0 & \text{if} \quad r|\mathbf{w}(\mathbf{P}) = r|\mathbf{w}(P_i',\mathbf{P}_{-i}).
\end{cases}
\end{equation}
%\Delta r|\mathbf{w}(\mathbf{P};{P'_i})
%where $\mathbf{P'}=(P_1,\dots,P_{i-1},P_i',P_{i+1},\dots,P_n)$. 
%
We stay agnostic about the precise source of perturbations: 
the switch from $P_i$ to $P_i'$ might reflect a spontaneous change of mind or intentional preference misrepresentation, e.g., because someone has bought $i$'s vote. 
Variations might also be the result of a mistake or of receiving last-minute private information about some of the candidates. 
Our important premise %of our quantification strategy 
is only that: a committee member's input to the collective decision process matters more, the more influential player~$i$ is in the committee and vice~versa. 

We can then quantify player~$i$'s a~priori influence or power -- and compare it to that of other players or for variations of $r|\mathbf{w}$ -- by taking expectations over all $(m!)^n$ conceivable $\mathbf{P}$ and all $m!-1$ possible perturbations of $P_i$ at any given $\mathbf{P}$: 
%,  sensitivity of outcome $r|\mathbf{w}(\mathbf{P})$ at profile $\mathbf{P}$ to a perturbation of $i$'s preferences is  $\frac{1}{(m!-1)} \sum_{P_i'\neq P_i \in \mathcal{P}(A)} \Delta r|\mathbf{w}(\mathbf{P};{P'_i})$. :  
%averaging the profile $\mathbf{P}$-specific outcome sensitivity across all $(m!)^n$ conceivable preference profiles $\mathbf{P}\in\mathcal{P}(A)^n$:
\begin{equation}\label{eq:expected_Delta}
\widehat{\mathcal{I}}_i(N,A,r|\mathbf{w}) := 
\mathbb{E}\big[\Delta r|\mathbf{w}(\mathbf{P};{P'_i})\big]=\frac{\sum_{\mathbf{P} \in \mathcal{P}(A)^n} \sum_{P_i'\neq P_i \in \mathcal{P}(A)} \Delta r|\mathbf{w}(\mathbf{P};{P'_i})}{(m!)^n\cdot (m!-1)}, \;\; i \in N.
\end{equation}
A value of $\widehat{\mathcal{I}}_i(N,A,r|\mathbf{w}) = 0.25$, for example, %$\widehat{\mathcal{I}}_i(r|\mathbf{w})=0.5$ 
%$\mathbb{E}\big[\Delta r|\mathbf{w}(\mathbf{P};{P'_i})\big]=0.5$ means 
means that 25\% of player $i$'s preference variations would change the outcome.

The expected value in (\ref{eq:expected_Delta}) equals the probability that a change of player $i$'s preferences from $P_i$ to random $P_i'\neq P_i$ at a randomly drawn profile $\mathbf{P}$ affects the outcome.
It is zero if and only if 
%$\widehat{\mathcal{I}}_i(r|\mathbf{w})$ is zero if 
player $i$ is a \emph{null player}, i.e., its preferences never make a difference to the committee decision. 
However, $\widehat{\mathcal{I}}_i(N,A,r|\mathbf{w})$ falls short of one for a \emph{dictator player}, i.e., when $r|\mathbf{w}(\mathbf{P})=a^*$ if and only if player~$i$ ranks $a^*$ top: 
since only changes of the dictator's top preference matter, only $(m!-(m-1)!)$ out of $m!-1$ perturbations of $P_i$ affect the outcome. 
We suggest to normalize power indications so that they range from zero to one. % (null to dictator player). 
Specifically, we %divide $\mathbb{E}\big[\Delta r|\mathbf{w}(\mathbf{P};{P'_i})\big]$ by $(m!-(m-1)!)/(m!-1)$ and 
focus on the voting power index $\mathcal{I}(\cdot )$ with %(N,A,r|\mathbf{w}
\begin{equation}\label{eq:I_definition}
\mathcal{I}_i(N,A,r|\mathbf{w}):=\frac{\mathbb{E}\big[\Delta r|\mathbf{w}(\mathbf{P};{P'_i})\big]}{(m!-(m-1)!)/(m!-1)}=
\frac{\sum_{\mathbf{P} \in \mathcal{P}(A)^n} \sum_{P_i'%\neq P_i 
		\in \mathcal{P}(A)} \Delta r|\mathbf{w}(\mathbf{P};{P'_i})}{(m!)^n\cdot(m!-(m-1)!)}
, \;\; i \in N,
\end{equation}
denoting \emph{player $i$'s a priori influence} or \emph{voting power} in weighted committee $(N,A,r|\mathbf{w})$. 

The normalization destroys $\widehat{\mathcal{I}}_i(N,A,r|\mathbf{w})$'s interpretation as a probability but facilitates comparison across committees. Regardless of how many alternatives and players are involved, % even when different numbers $m$ of alternatives and numbers $n$ of players are involved:
$\mathcal{I}_i(N,A,r|\mathbf{w})\in[0,1]$ indicates how close player~$i$ is to being a dictator in $(N,A,r|\mathbf{w})$. % and how far from being a null player.
$\mathcal{I}_i(N,A,r|\mathbf{w})=0.5$, for instance, implies that $i$'s influence lies halfway between that of a null player and a dictator. So, on average, outcomes are half as sensitive to $i$'s preferences than they would be if $i$ commanded all votes. 

\subsection{Relationship to the Penrose-Banzhaf Index\label{sec:PBI}}

For $m=2$ alternatives, the denominators in (\ref{eq:expected_Delta}) and (\ref{eq:I_definition}) equal $2^n$ and ${\mathcal{I}}_i(N,A,r|\mathbf{w})=\widehat{\mathcal{I}}_i(N,A,r|\mathbf{w})$. Moreover,  for any of the rules~$r$ introduced above,
%in Section~\ref{sec:rules}, 
weighted committees coincide with the subclass of {simple voting games} $(N,v)$ where $v(S)\in \{0,1\}$ is given by $v(S)=1 \Leftrightarrow w(S)\ge \frac{1}{2}w(N)$
with $w(T):=\sum_{i\in T}w_i$ for all $T\subseteq N$.
%We can refer to $(N,v)$ as 
If we consider the simple game $(N,v)$ induced by $(N,\{a_1,a_2\},r|\mathbf{w})$,
its Penrose-Banzhaf index $PBI(N,v)$ turns out to coincide with %voting power index 
${\mathcal{I}}(N,A,r|\mathbf{w})$.
Namely, $PBI(\cdot)$'s usual definition then specializes to
\begin{align}
PBI_i(N,v)&:=\frac{1}{2^{n-1}} \sum_{S\subseteq N\setminus \{i\}} [v(S \cup \{i\})-v(S)]  \\ 
%= \frac{1}{2^{n-1}} \sum_{S\subseteq N, i\in S} [v(S)-v(S\setminus \{i\})]
& =\frac{\big|\{S\subseteq N\setminus \{i\}: w(S)< \frac{1}{2}w(N)%\} \cap \{S\subseteq N\setminus \{i\}:
	\text{\ \  and\ \ } w(S\cup\{i\})\ge \frac{1}{2}w(N)\}\big|}{2^{n-1}}. \notag
\end{align}
%when $(N,v)$ is induced by $(N,\{a_1,a_2\},r|\mathbf{w})$. % for $v(S)=1 \Leftrightarrow w(S)\ge \frac{1}{2}w(N)$. 
In this case, for $r\in \{r^B, r^C, r^P, r^{PR}, r^S\}$ with lexicographic tie-breaking,
\begin{equation}
\begin{array}{rcr}
\Delta r|\mathbf{w}(\mathbf{P};{P'_i})=1 \quad 
&  \Leftrightarrow & [ r|\mathbf{w}(P_i, \mathbf{P}_{-i})=a_1 \text{\ \ and \ \,} r|\mathbf{w}(P_i', \mathbf{P}_{-i})=a_2 ] \\
& & \text{ or \ \  } [ r|\mathbf{w}(P_i, \mathbf{P}_{-i})=a_2 \text{\  \ and \ \,} r|\mathbf{w}(P_i', \mathbf{P}_{-i})=a_1 ] \\
&  \Leftrightarrow &  w(\overbrace{\{j\neq i\in N: a_1 P_j a_2 \}}^{S^{\mathbf{P}_{-i}}}) + w_i \ge \frac{1}{2}w(N) \\
& & \text{ and \ \ } w(\{j\neq i\in N: a_2 P_j a_1 \})+ w_i > \frac{1}{2}w(N) \medskip \\
&  \Leftrightarrow & w(S^{\mathbf{P}_{-i}})< \frac{1}{2}w(N)
\text{\ \ and \ }  w(S^{\mathbf{P}_{-i}}\cup \{i\}) \ge \frac{1}{2}w(N). 
\end{array}
\end{equation}
The last line substitutes
$w(\{j\neq i\in N: a_2 P_j a_1 \})=w(N)- w_i - w(S^{\mathbf{P}_{-i}})$ where $S^{\mathbf{P}_{-i}}:=\{j\neq i\in N: a_1 P_j a_2 \}$. % and $\bar S^{\mathbf{P}_{-i}}=\{j\neq i\in N: a_2 P_j a_1 \}$. 
Hence
\begin{align}
\mathcal{I}_i(N,\{a_1,a_2\},r|\mathbf{w}) 
&=\frac{\sum_{\mathbf{P} \in \mathcal{P}(A)^n} \sum_{P_i' \in \mathcal{P}(A)} \Delta r|\mathbf{w}(\mathbf{P};{P'_i})}{2^n} \\
&= \frac{2\cdot \big|\{S\subseteq N\setminus \{i\}: w(S)< \frac{1}{2}w(N) %\} \cap \{S\subseteq N\setminus \{i\}: 
	\text{\ \ and\ \ }w(S\cup\{i\})\ge \frac{1}{2}w(N)\}\big|}{2^n} 
%\\ & 
= PBI(N,v) \notag
\end{align}
as every $S=S^{\mathbf{P}_{-i}}\subseteq N\setminus \{i\}$ arises for two  profiles $(P_i,\mathbf{P}_{-i})\in \mathcal{P}(A)^n$, one involving $a_1 P_i a_2$ and the other $a_2 P_i a_1$.

\section{Illustration} \label{sec:illustration}
\subsection{A Toy Example\label{sec:example}}
As a first illustration let us evaluate the distribution of voting power when our stylized hiring committee %(Section~\ref{sec:Introduction}) 
(see Introduction) with three groups of 6, 5, and 3 members adopts Borda rule $r^B$, that is weighted committee $(N,A,r^B|(6,5,3))$. 
With $|A|=2$ candidates, the applicant ranked first by any two groups wins. 
So all three players are symmetric and have identical power according to the Penrose-Banzhaf or any other established voting power index. 

The symmetry is broken when three or more candidates are involved. Given $A=\{a,b,c\}$, $\mathcal{I}(N,A,r^B|(6,5,3))$ 
evaluates all $(3!)^{3}=216$ strict preference profiles $\mathbf{P}\in \mathcal{P}(A)^3$ and checks whether a change of player~$i$'s respective preference $P_i$ makes a difference to the Borda winner. 
Table~\ref{table:big_Borda_table} illustrates this for profile $\mathbf{P}=(bca, abc, cba)$. 
The Borda winner $b$ at $\mathbf{P}$ %$r^B|(6,5,3)(\mathbf{P})=b$ 
has a score of $20 =6\cdot 2+5\cdot 1+3\cdot 1$ points vs.\ 10 for $a$ vs.\ 12 for $c$ (first block of table). 
When preferences $P_1=bca$ of group~1 are varied (second block), changes to $P_1'\in \{abc, acb, cab, cba\}$ result in a new Borda winner (indicated by an asterisk) while $P_1'=bac$ does not. 
Similarly, three out of five perturbations $P_2'$ of player~2's preferences %$P_2=abc$ 
would change the outcome (third block); however,
no variation of $P_3$ %=cba$ 
affects the committee choice (last block). The considered profile $\mathbf{P}=(bca, abc, cba)$ therefore contributes $(\sfrac{4}{864}, \sfrac{3}{864},0)$ to $\mathcal{I}(\cdot)$. %N,A,r^B|(6,5,3))$.

\begin{table}[htbp]
\begin{tabular}{||ccc||c|c|c|c||c|c|c|c||c|c|c|c||}
	\hline\hline
	$      $ & $\mathbf{P}=$ &        & $P_1'$ &      $a$       &      $b$      &      $c$       & $P_2'$ & $a$ &      $b$      &      $c$       &     $P_3'$     &      $a$      &      $b$       &  $c$  \\
	\hline
	$(bca,$  & $abc,$        & $cba)$ & $abc$  & $\mathbf{22}$* &      14       &       6        &   -    &  -  &  -            &  -             &     $abc$ & 16 & $\mathbf{20}$ & 6  \\
	         & $\Downarrow$  &        & $acb$  & $\mathbf{22}$* &       8       &       12       & $acb$  & 10  &      15       & $\mathbf{17}$* &     $acb$ & 16 & $\mathbf{17}$ & 9  \\
	$a$      & $b$           & $c$    & $bac$  &       16       & $\mathbf{20}$ &       6        & $bac$  &  5  & $\mathbf{25}$ &       12       &     $bac$ & 13 & $\mathbf{23}$ & 6  \\
	10       & $\mathbf{20}$ & 12     &   -    &        -       &      -        &      -         & $bca$  &  0  & $\mathbf{25}$ &       17       &     $bca$ & 10 & $\mathbf{23}$ & 9  \\
	         &               &        & $cab$  &       16       &       8       & $\mathbf{18}$* & $cab$  &  5  &      15       & $\mathbf{22}$* &     $cab$ & 13 & $\mathbf{17}$ & 12 \\
	         &               &        & $cba$  &       10       &      14       & $\mathbf{18}$* & $cba$  &  0  &      20       & $\mathbf{22}$* &       -   &  - &    -          &   -\\
	\hline\hline
\end{tabular}
	\caption{\small Effect of perturbation of $\mathbf{P}=(bca, abc, cba)$ to $(P_i',\mathbf{P}_{-i}$) on Borda scores}
	\label{table:big_Borda_table}
\end{table}
Aggregating the corresponding figures for all $\mathbf{P}\in \mathcal{P}(A)^n$ yields 
\begin{equation}\label{eq:example_I}
\mathcal{I}(N,A,r^B|(6,5,3))=(\sfrac{588}{864}, \sfrac{516}{864},\sfrac{312}{864})\approx(0.6806, 0.5972, 0.3611).
\end{equation} 
So for the weights at hand, group~1 has almost 70\% of the opportunities to swing the collective choice that it would have when deciding alone. 
This figure is roughly halved for group~3 even though both are symmetric when choosing between two candidates. 
The comparison shows that traditional voting power indices for simple voting games, such as $PBI(\cdot)$, can yield highly misleading conclusions when decisions of the investigated institution involve more than $m=2$ alternatives.
(This is one of the shortcomings alluded to earlier.)
One can also see from the numbers in (\ref{eq:example_I}) that $\mathcal{I}(\cdot)$ need not add to one: the collective outcome at a given $\mathbf{P}$ may be sensitive to the preferences of several players at the same time, or to those of none.\footnote{Therefore, %the 
normalization $\overline{PBI}(N,v)=PBI(N,v)/\sum_i PBI_i(N,v)$ is sometimes %invoked 
applied in binary voting analysis.}

\begin{table}[htbp] 
	\small
	\centering
	\begin{tabular}{||c|c|c|c||}
		\hline\hline
		&           $m=3$            &          $m=4$           &          $m=5$           \\
		\hline
		$\mathcal{I}(r^P|(6,5,3))$    &   (0.6667, 0.4444, 0.4444)   & (0.7500, 0.3750, 0.3750) & (0.8000, 0.3200, 0.3200) \\
		\hline
		$\mathcal{I}(r^{PR}|(6,5,3))$ &  (0.5556, 0.5556, 0.5000)  & (0.5833, 0.5833, 0.5000) & (0.6000, 0.6000, 0.5000) \\
		\hline
		$\mathcal{I}(r^{B}|(6,5,3))$  & (0.6806, 0.5972, 0.3611) & (0.7372, 0.6246, 0.3644) & (0.7631, 0.6462, 0.3839) \\
		\hline
		$\mathcal{I}(r^{C}|(6,5,3))$  &  (0.5509, 0.5509, 0.5509)   & (0.5851, 0.5851, 0.5851) & (0.6098, 0.6098, 0.6098) \\
		\hline
		$\mathcal{I}(r^{S}|(6,5,3))$  &  (0.5972, 0.5278, 0.5278)  & (0.6584, 0.5426, 0.5426) & (0.7011, 0.5515, 0.5515) \\
		\hline\hline
	\end{tabular}
	\caption{\small Voting power in committee $(N, A, r|(6,5,3))$ for $|A|=m$ and $r\in\{r^P,r^{PR},r^B,r^C,r^S\}$}
	\label{table:results_example} 
\end{table}

Of course, manual computations as in Table~\ref{table:big_Borda_table} are tedious. %manually. 
It is not difficult, though, to evaluate  $\mathcal{I}(\cdot)$  with a standard desktop computer for up to five alternatives;  
and to compare the respective distribution of voting power to that arising from other voting rules. 
Findings are summarized for $r\in \{r^P,r^{PR},r^B,r^C,r^S\}$ in Table~\ref{table:results_example}. 
As the comparison between $m=2$ and 3 showed already, voting powers vary in the number of alternatives. % faced by the committee. 
Under plurality rule, for instance, player~1 is closer to having dictatorial influence, the more alternatives split the vote of players~2 and 3. 
$\mathcal{I}(N,\{a_1,\ldots, a_m\},r^P|(6,5,3))$ tends to $(1,0,0)$ as $m\to \infty$ (given sincere voting and independent preferences).%
\footnote{Comparative statics are more involved for other rules: bigger $m$ tends to raise the share of profiles $\mathbf{P}$ at which \emph{some} perturbation of $P_i$ affects the outcome but lowers the fraction of perturbations $P_i'$ that do so for both player~$i$ and the hypothetical dictator used as normalization. 
Superposition of these effects here increases power for all players and $r\in \{r^{PR},r^B,r^C,r^S\}$, but this is not true in general.}
That power of all three players coincides under $r^C$ %follows from the observation
confirms that %$r^C$ 
Copeland rule extends the symmetry relation between players for $m=2$ alternatives to any number $m>2$ (see \shortciteNP[Prop.~3]{Kurz/Mayer/Napel:2018}).

\subsection{Election of the IMF's Managing Director\label{sec:IMF}}

Evaluation of $\mathcal{I}(\cdot)$ is, of course, more interesting for real-world institutions than toy examples. We  pick the International Monetary Fund  (IMF) as a case in point. 
Binary power indices have been applied to it many times. See, e.g., Leech~\citeyear{Leech:2002,Leech:2003}, \citeN{Aleskerov/Kalyagin/Pogorelskiy:2008}, or \citeN{Leech/Leech:2013}. 
We extend the analysis to election of the IMF's Managing Director from three candidates by the Executive Board.

The Executive Board consists of 24 members whose voting weights reflect financial contributions to the IMF, so-called \emph{quotas}. 
The six largest contributors -- USA, China, Japan, Germany, France and the UK -- and Saudi Arabia currently provide one Executive Director each. The remaining 182 countries are grouped into seventeen constituencies. 
Each supplies one Executive Director who represents the group members and wields their combined voting rights. 

Various changes to the distribution of quotas have taken place since the IMF's inception in 1944 at Bretton Woods. The most recent reform was agreed in 2010 and started to be implemented in 2016. 
A significant share of votes has shifted from the USA and Europe to emerging and developing countries.
%\footnote{The voting share of the USA still dwarfs all other members. The USA were nevertheless one of the last countries to ratify the reform.} 
In particular, China's vote share has gone up to 6.1\% (compared to 3.8\% before). India's share increased to 2.6\% (2.3\%), Russia's to 2.6\% (2.4\%), Brazil's to 2.2\% (1.7\%) and Mexico's to 1.8\% (1.5\%). 
%This brings India and Brazil into the list of the top ten IMF members in terms of vote shares.
On the occasion,
%In addition to reforming voting weights, 
the IMF also modified the election process for its key representative, the Managing Director (currently: Christine Lagarde). 

Prior to the reform, the process was criticized as intransparent and undemocratic: 
the Managing Director used to be a European chosen in backroom negotiations with the US. 
The new process is advertised as ``open, merit based, and transparent'' (IMF Press Release 16/19): 
all Executive Directors and IMF Governors may nominate candidates. 
If the number of nominees is too big, a shortlist of three candidates is drawn up based on indications of support. 
From this shortlist the new Managing Director is elected ``by a majority of the votes cast'' in the Executive Board.\footnote{IMF Press Release 16/19, Part~4, holds that ``Although the Executive Board may select a Managing Director by a majority of the votes cast, the objective of the Executive Board is to select the Managing Director \mbox{by consensus \ldots''.} The same is said in Part~3 about adoption of the ``shortlist''. Our analysis presumes that a consensus may not always be found -- or it actually arises in the shadow of the anticipated outcome of %weighted 
voting.}

\renewcommand{\arraystretch}{1.1}
\begin{table}
	\centering
	\begin{tabular}{||l|c|c|c|c|c|c|c|c||}
		\hline \hline 
		& \multicolumn{2}{c|}{Vote share (\%)}  & \multicolumn{2}{c|}{$\mathcal{I}(r^P|\mathbf{w})$} & \multicolumn{2}{c|}{$\mathcal{I}(r^{PR}|\mathbf{w})$} & \multicolumn{2}{c||}{$\mathcal{I}(r^C|\mathbf{w})$} \\
		\cline{2-9}
		&$\mathbf{w}_{\text{pre}}$ &$\mathbf{w}_{\text{post}}$   & $\mathbf{w}_{\text{pre}}$ & $\mathbf{w}_{\text{post}}$  & $\mathbf{w}_{\text{pre}}$  & $\mathbf{w}_{\text{post}}$ & $\mathbf{w}_{\text{pre}}$  & $\mathbf{w}_{\text{post}}$  \\
		\hline   
		USA    & 16.72  & 16.47& 0.7126  &0.7030 &0.6740  &0.6653 & 0.6880  &0.6790 \\
		Japan &  6.22   & 6.13&0.1986   & 0.1989  &0.2239 & 0.2233& 0.2164 &0.2159 \\
		China  & 3.80   & 6.07&0.1216   &0.1967   &0.1404 &0.2209 &0.1340  &0.2135 \\
		\emph{Netherlands} &  6.56   & 5.41 &0.2092  &0.1755  &0.2350 &0.1983 &0.2277  &0.1910 \\
		Germany &  5.80  &5.31&0.1851 & 0.1720  &0.2097  & 0.1950&0.2024 & 0.1876\\
		\emph{Spain} &  4.90  &5.29 & 0.1567 &0.1718  &0.1789  &0.1945 & 0.1717  & 0.1871\\
		\emph{Indonesia} &  3.93  &4.33&0.1254  &0.1403  &0.1448  &0.1607 &0.1382  &0.1538 \\
		\emph{Italy} &  4.22   &4.12&0.1349  &0.1337  &0.1551 & 0.1533& 0.1482  & 0.1465\\
		France &   4.28   &4.02&0.1370  &0.1306   & 0.1574 &0.1499 &0.1507 &0.1432 \\
		United Kingdom &  4.28   &4.02&0.1369  &0.1304 &0.1574  &0.1498 &0.1506  &0.1431 \\
		\emph{Korea} &  3.48   &3.78&0.1114  &0.1226 & 0.1291  & 0.1410& 0.1230 &0.1345 \\
		\emph{Canada} &  3.59   &3.37& 0.1150  &0.1093 &0.1332  &0.1265 &0.1268 & 0.1203\\
		\emph{Sweden} &   3.39   &3.28& 0.1085 & 0.1063  & 0.1259 &0.1231 & 0.1198 & 0.1171\\
		\emph{Turkey} &  2.91  & 3.22& 0.0932  & 0.1044  & 0.1088 & 0.1209& 0.1032 & 0.1149\\
		\emph{South Africa} & 3.41  &3.09& 0.1091  & 0.1001  &0.1267  &0.1162 &0.1205 &0.1104 \\
		\emph{Brazil} &  2.61   &3.06&0.0835 &0.0993  &0.0979  &0.1154 & 0.0927 & 0.1096\\
		\emph{India} & 2.80    &3.04&0.0898  &0.0988 &0.1048 & 0.1147& 0.0993 & 0.1089\\
		\emph{Switzerland} &   2.94   &2.88&0.0941  &0.0935   &0.1097  &0.1087 &0.1041 &0.1030 \\
		\emph{Russian Federation} & 2.55   &2.83&0.0817   & 0.0920 &0.0957  &0.1070 &0.0905 &0.1015 \\
		\emph{Iran} &  2.73   &2.54& 0.0874  &0.0823   & 0.1024  &0.0962 & 0.0970 & 0.0910\\
		\emph{Utd.~Arab Emirates} & 2.57   &2.52& 0.0822   &0.0817 &0.0963  &0.0955 & 0.0911 & 0.0904\\
		Saudi Arabia &  2.80   &2.01& 0.0896  &0.0652  & 0.1046  & 0.0767&0.0992  & 0.0723\\
		\emph{Dem.~Rep. Congo} & 1.46  &1.62& 0.0465  &0.0526  &0.0555 &0.0621 & 0.0521 & 0.0584\\
		\emph{Argentina} &  1.84   &1.59& 0.0587 &0.0515  &0.0695 &0.0610 &0.0654 & 0.0573\\
		\hline
		\hline 
	\end{tabular}
	\caption{\small Influence %$\mathcal{I}(N,A,r|\mathbf{w})$ 
		in IMF Executive Board for pre- and post-reform weights and $m=3$ \\ 
		% $\mathbf{w}_{\text{pre}}$) and after implementation of 2016 reform ($\mathbf{w}_{\text{post}}$); 
		(\emph{groups} as of Dec.~2018 indicated by largest member, Nauru included in $\mathbf{w}_{\text{post}}$) %; pre-reform scenario does not include Nauru, which joined the IMF in April 2016. 
		%as its 189th member and belongs to the Korean group.
	} 
	\label{table:IMF_results}
\end{table}

The IMF has neither publicly nor upon our email request specified what ``majority of the votes cast'' exactly means to it for three candidates. 
We take the resulting room for interpretation as an opportunity to simultaneously investigate the voting power effects of the weight reform and of a procedural choice between using $(i)$~plurality rule, $(ii)$~plurality with a runoff if none of three shortlisted candidates secures 50\% of the votes, or $(iii)$~Copeland rule. 
In the spirit of earlier a~priori analysis of the IMF, we maintain the independence assumption that underlies $PBI(\cdot)$ and $\mathcal{I}(\cdot)$.
This provides an a~priori assessment of how level is the playing field created by weights as such rather than an estimate of who wields how much influence on the next decision given prevailing political ties, economic interdependencies, etc.

Influence figures in Table~\ref{table:IMF_results} are based on Monte Carlo simulations with sufficiently many iterations so that differences within rows are significant with $\ge\! 95\%$ confidence.\footnote{ 
The only exception is that the difference between $\mathcal{I}_{\text{Japan}}(r^P|\mathbf{w_{\text{pre}}})$ and $\mathcal{I}_{\text{Japan}}(r^P|\mathbf{w_{\text{post}}})$ is not signficant. The large number $6^{24}>4.7\cdot 10^{18}$ of preference profiles renders exact calculation of $\mathcal{I}(\cdot)$ 
impractical.} % compared to Monte Carlo simulation.}  
We find that 2016's increase of vote shares for emerging market economies has indeed raised their voting power,  %translate into more influence in the Executive Board.\footnote{We remark that this also holds for $m=2$.} 
no matter which voting rule %$(i)$--$(iii)$ 
we consider. 
This is most pronounced for China, with an increase of more than 50\%. 
%Other emerging market economies also have a greater say than before the 2016 reform. The i
Influence of the groups led by Brazil and Russia (incl.\ Syria) increased by about 18\% and 12\%, respectively; that of the Turkish and Indonesian group by about 11\% each; 
the Indian and Spanish (incl.\ Mexico and others) groups gained about 10\% and 9\%, respectively. 
Intended or not, %many African countries are relatively big losers of voting power: 
the South African group lost about 8\% of its a~priori voting power; Saudi Arabia is the greatest loser with roughly 27\%. 
Germany, France and UK each lost between 5\% and 7\% while voting power of the USA  stayed largely constant. %unaffected by the weight reform. % and lose only about 1\% of their influence.

The computations exhibit %Table~\ref{table:IMF_results} 
%shows whether adopting one or the other of the possible interpretations of the selection process for the Managing Director makes a difference. It turns out that they really differ for countries' chances to influence the outcome. There is 
a simple pattern regarding the adopted interpretation of `majority': % how ``by a majority of the votes cast'': while 
voting power of the USA is higher for plurality rule than for Copeland than for plurality runoff; the opposite applies to all other (groups of) countries. %They all have more influence under plurality runoff rule than under Copeland rule than under plurality rule. This is in line with 
This echoes findings for our toy example: %in Section~\ref{sec:example}: 
the largest player's influence was highest for $r^P$; small and medium players were more influential under $r^{PR}$ and $r^C$.

%the estimated influence numbers; respective 95~\%-confidence intervals are provided in Appendix~A (Tables~\ref{table:IMF_confidence1} to \ref{table:IMF_confidence3}). 
%Auf Nassau und Sudan hinweisen

\section{Towards More General Rule Comparisons\label{sec:general_comparison}}

It seems worthwhile to check whether %such casual 
observations like the ones above are robust: does the largest group benefit from plurality votes or the smallest from pairwise comparisons in general? 
And can recommendations for maximizing a player's influence be given also if information about the exact distribution of voting weights is vague or fluctuates? 
We take a first step beyond specific examples and look for possible size biases of the rules.
Attention is restricted to small numbers of players and alternatives, namely $n=m=3$. The respective intuitions may still apply more generally to shareholder meetings, weighted voting in political bodies, etc.

%We take up the geometric approach of \shortciteN{Kurz/Mayer/Napel:2018} %, where all structurally distinct weighted committee games for $n=3$ players were illustrated.
%In order to represent %summarize voting rules %, two at a time, 
%%for 
%all possible weight distributions among three players, 
We use the standard projection of the 3-dimensional simplex of relative weights to the plane in order to represent all possible weight distributions among three players: vertices give 100\% of voting weight to the indicated player, the midpoint corresponds to $(\sfrac{1}{3},\sfrac{1}{3},\sfrac{1}{3})$, etc.  %(Figure~\ref{fig:simplex}). %The weight axes are suppressed in subsequent figures.
Each figure %(also see %Figures~\ref{fig:Borda_Plurality_3_3}--\ref{fig:PluralityRunoff_Schulze_3_3} in 
%the Appendix) 
presents the result of comparing influence of player~1 under some rule~$\rho_A$ vs.\ rule~$\rho_B$. Areas colored green (red) indicate weight distributions for which 
$\mathcal{I}_1(N,A, \rho_A) >\!\!\mbox{$(<)$}\ \mathcal{I}_1(N,A, \rho_B)$.
%$\mathcal{I}_1(\{1,2,3\},\{a,b,c\}, \rho_A) > \mathcal{I}_1(\{1,2,3\},\{a,b,c\}, \rho_B)$
Borda rule was found to be particularly sensitive to weight differences by \shortciteN{Kurz/Mayer/Napel:2018}; when it is involved in a comparison, we use darker tones of green or red to indicate greater influence differences. 

%\begin{figure}
%	\begin{center}
%	\includegraphics[width=0.69\textwidth]{Borda_Plurality_3_3_example.pdf}
%	\caption{\small Borda vs. plurality for %$n=3$ and 
%		$m=3$. Regions colored green (yellow/red) indicate that Borda influence is greater than (equal to/smaller than) plurality influence\label{fig:Borda_Plurality_3_3_example}}
%\end{center}
%\end{figure}

\subsection{Borda vs.\ Plurality}
\begin{figure}[h!]
	\begin{center}
	\includegraphics[width=0.69\textwidth]{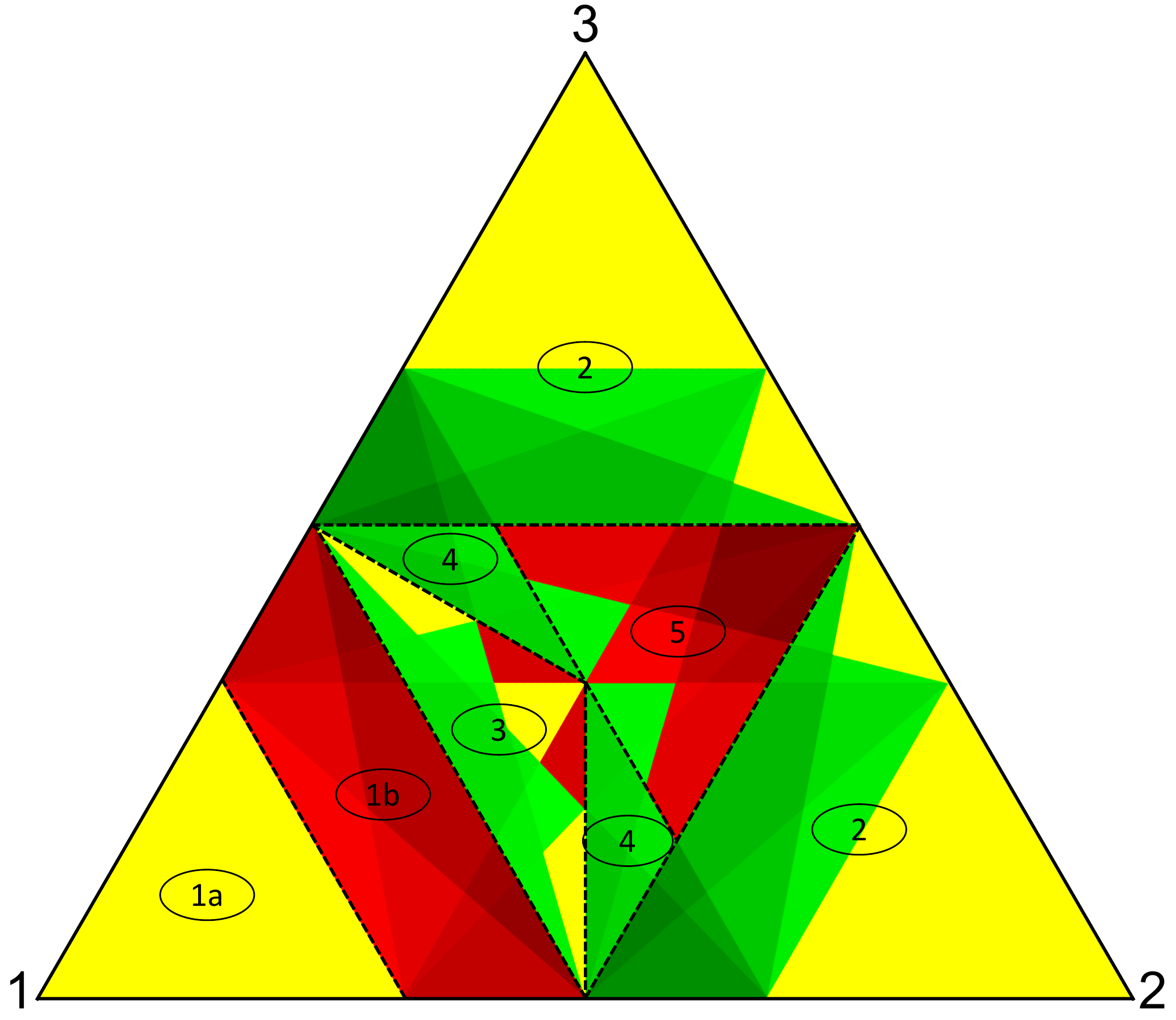}
	\caption{\small Borda vs. plurality for %$n=3$ and 
		$m=3$. Regions colored green (yellow/red) indicate that Borda influence is greater than (equal to/smaller than) plurality influence\label{fig:Borda_Plurality_3_3_example}}
\end{center}
\end{figure}

The major cases that arise when we compare player~1's influence in Borda vs.\ plurality committees are numbered in Figure~\ref{fig:Borda_Plurality_3_3_example}. %\footnote{The figure is also included without annotations as Figure~\ref{fig:Borda_Plurality_3_3} in the Appendix.} 
We focus on $w_1 \neq w_2 \neq w_3$ and write $\tilde w_i=w_i/(w_1+w_2+w_3)$, $ w_{-1}^+=\max \{\tilde w_2, \tilde w_3\}$, and $ w_{-1}^-=\min \{\tilde w_2, \tilde w_3\}$. 
The following recommendations could be given to an influence-maximizing player~1 if
the procedural choice between $r^B$ and $r^P$ is at this player's discretion: % then the following ``rules of thumb'' could be used:

\begin{itemize}
	\item If you wield the majority of votes (region 1) impose plurality rule.  \vspace{0.1cm} \\
	Namely, $\tilde w_1>\frac{2}{3}$ makes you both a plurality and Borda dictator (region 1a); 
	$\frac{2}{3}\ge\tilde{w_1}>\frac{1}{2} $ implies dictatorship only under plurality rule (region 1b). 
	\item Also impose plurality rule (region~5) \vspace{-0.3cm}
	\begin{itemize}
\item if your weight is smallest and both other players have a third to half of the votes each ($\frac{1}{3}\le w_{-1}^-<\frac{1}{2}$), or 
\item if you have less than a third of votes and the largest player falls short of the majority by no more than a quarter of the remaining player's votes 
	($\frac{1}{2}>w_{-1}^+\ge \frac{1}{2} -\frac{1}{4}w_{-1}^-$).
\end{itemize}
\vspace{-0.4cm}
	\item Otherwise, as a good `rule of thumb',  impose Borda rule. \vspace{0.1cm} \\
	Namely, when some other player holds the majority (region~2, $w_{-1}^+>\frac{1}{2}$) the observations for region~1 essentially get reversed. 
	In case that nobody holds the majority, Borda comes with greater influence if you are second-largest with at least a third of votes (region~4, $\frac{1}{2}>w_{-1}^+>\tilde{w_1}\ge \frac{1}{3}$). This extends weakly to when you are largest (region~3). The only exception to the rule are two small subregions where all weights are similar but $\tilde w_1\ > \tilde w_{-1}^+ >\frac{1}{3}>\tilde w_{-1}^-$. % and $r^P$ gives player~1 slightly higher influence than $r^B$.
\end{itemize}

\subsection{Further Comparisons}
Analogous pairwise influence comparisons are depicted in the Appendix for all possible weight distributions $\mathbf{w}$ and $r\in \{r^B, r^C, r^P, r^{PR}, r^S\}$. Again, Borda's high responsiveness to weight differences makes for more detailed case distinctions (see Figures~\ref{fig:Borda_Plurality_3_3}--\ref{fig:Borda_Schulze_3_3}). 
By contrast,
when plurality rule is compared to either Copeland, plurality runoff, or Schulze rule (Figure~\ref{fig:Plurality_PluralityRunoffCopelandSchulze_3_3}), the recommendation always is simple and intuitive: plurality rule maximizes influence if you have the most votes. If anyone else has more votes, your influence is greater (at least weakly) under the respective other rule. %you are either equally influential, or the respective other rule makes you more influential).

%
%
%Similar reasoning also applies to all other comparisons. The corresponding figures are provided in Appendix~B. For comparisons not involving Borda rule, things are rather simple. For instance, when plurality rule is compared to either Copeland, plurality runoff, or Schulze rule (Figure~\ref{fig:Plurality_PluralityRunoffCopelandSchulze_3_3}), the recommendation always is the same: you should impose plurality rule if you have the most votes. If someone else has the most votes, you are better off under the respective other rule (you are either equally influential, or the respective other rule makes you more influential). %The only difference applies to cases in which you share a plurality with someone else. If the smallest player has a positive share of votes, then plurality is better than Copeland and equal to plurality runoff, but worse than Schulze. If two players each have exactly... 

Recommendations to an influence-maximizing player are similar for Copeland vs.\ Schulze rule (Figure~\ref{fig:Copeland_Schulze_3_3}): 
if you wield a plurality of votes, Schulze comes with greater influence; in case someone else has more votes, it is the opposite. 
For Copeland vs.\ plurality runoff (Figure~\ref{fig:Copeland_PluralityRunoff_3_3}), the former gives greater influence to you if you have at least the second-most votes. %; otherwise go for Copeland rule. 
If player~1 is to choose between plurality runoff and Schulze rule (Figure~\ref{fig:PluralityRunoff_Schulze_3_3}), Schulze rule gives greater influence if $w_1$ is either largest or smallest; otherwise it is better to adopt plurality runoff.

%Hier dann nochmal auf die Klassen eingehen und bei Plurality vs Copeland z.B. sagen, dass es kein Wunder ist, dass alles innerhalb des grossen gelben Dreiecks links unten aequivalent ist: Spieler 1 ist unter beiden Regeln Diktator. Das gleiche gilt fuer die Aequivalenz der anderen beiden grossen Dreiecke. Einmal ist Spieler 2 Diktator, ein anderes mal Spieler 3. Aus Sicht von Spieler 1 ist das aber egal, da er in beiden Fällen nichts zu sagen hat.

\subsection{Influence-maximizing Voting Rules}

\begin{figure}
	\centering
	\includegraphics[width=0.69\textwidth]{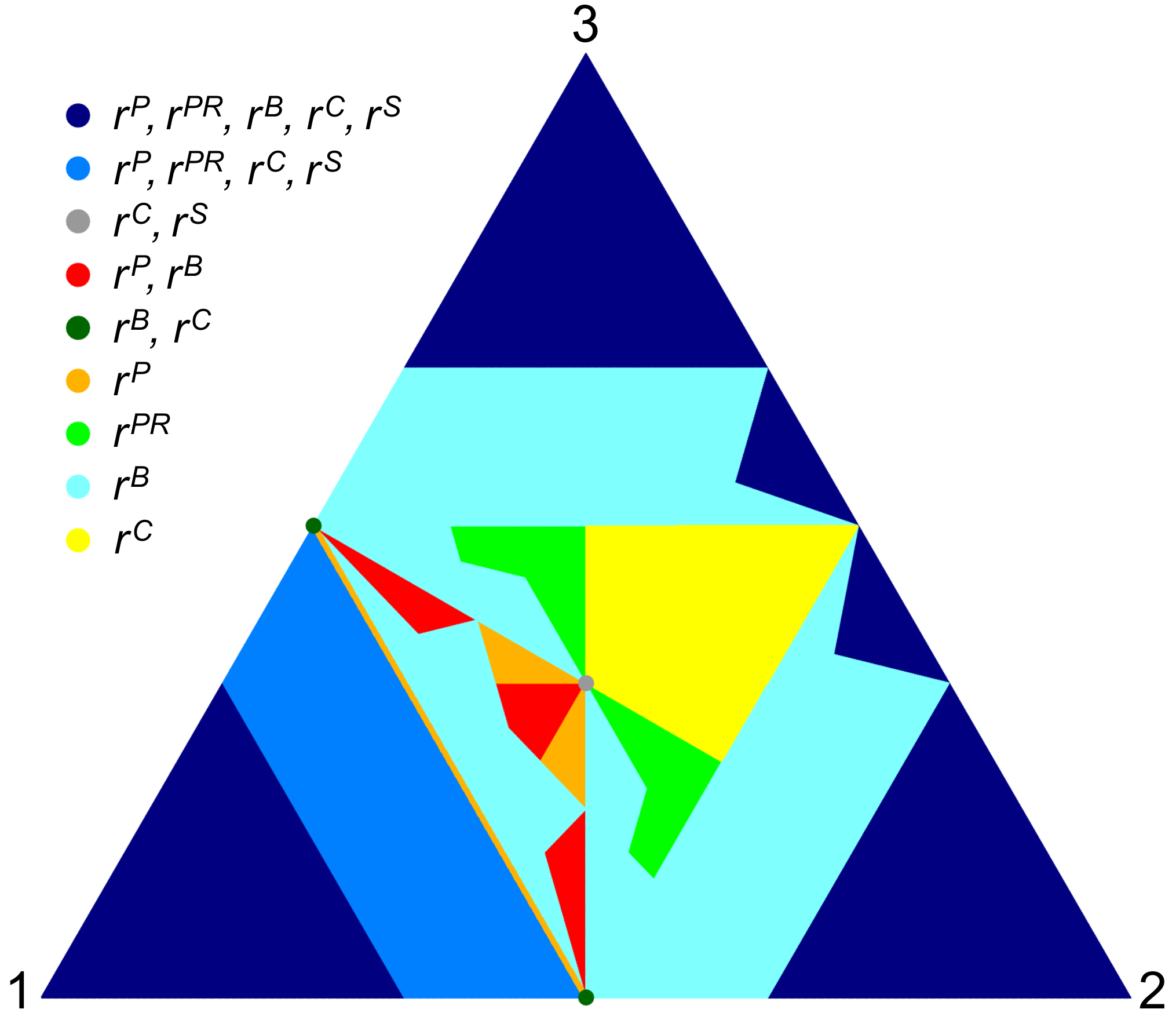}
	\caption{\small Maximal influence map %of the influence maximizing voting rules 
		for Borda, Copeland, plurality, plurality runoff, and Schulze rules for $m=3$.\label{fig:Best_rule_3_3}}
\end{figure}

Complementing pairwise comparisons as in Figures~\ref{fig:Borda_Plurality_3_3_example} and \ref{fig:Borda_Plurality_3_3}--\ref{fig:PluralityRunoff_Schulze_3_3}, one can also check directly which of the considered voting rules maximizes a specific player's a~priori voting power at any given weight distribution. This is done in Figure~\ref{fig:Best_rule_3_3}: 
configurations of same color indicate the same set of influence-maximizing voting rules for player~1.  % for all different distributions of voting weights. 
(When the respective weight regions are lines %segments 
or single points, we have manually enlarged them.) 
Tongue-in-cheek, Figure~\ref{fig:Best_rule_3_3} %Figure~\ref{fig:Best_rule_3_3} 
provides a `map' for influence-maximizing chairpersons -- or for whoever %decides on voting procedures and 
has a say on the adopted voting rule and cares about a~priori influence on decisions between three candidates.
It also gives players~2 or 3 reason for criticizing adoption of a particular rule as biased.

\section{Concluding Remarks}
%The figures, calculations and analysis above 
We have investigated how the distribution of %a~priori influence 
voting power in a committee depends not only on voting weights but which of the many possible aggregation procedures 
for $m>2$ alternatives is adopted -- from simple plurality voting to the elaborate computation of Schulze winners. 
Traditional measures of voting power, such as the Penrose-Banzhaf or Shapley-Shubik indices, fail to capture this. 
They should be accompanied by power indices for three or more alternatives. % (unless decision are purely binary, of course). 

One such index has been proposed and illustrated here. 
It evaluates how sensitive the collective choice under the given aggregation rule and weights is to preference changes by an individual player. 
The index is proportional to the probability for a random individual preference perturbation to affect the outcome, assuming preferences are independently uniformly distributed a priori.\footnote{It as an open challenge to find properties or `axioms', like those used by \citeN{Dubey/Shapley:1979} or \citeN{Laruelle/Valenciano:2001} for the Penrose-Banzhaf index, that would characterize the proposed index without making specific probability assumptions. Our attempts have failed so far.}
 A dictator player's voting power is normalized to one; a null player's power is zero.
The extent to which the distribution of weights matters %has been shown to 
differs across rules. So do comparative statics regarding the number of alternatives. 
How the adopted rule affects the distribution of voting power has been illustrated for (non-consensual) election of the IMF's Managing Director by its Executive Board. 
%influence  what individual members of the IMF's Executive Board can achieve if, e.g., it is to decide on three candidates for its next Managing Director. 
Similar analysis might be conducted for multi-candidate election primaries, party conventions, corporate boards, etc. 

Several case studies have shown with the benefit of hindsight how choice of a particular voting method may have affected big political decisions (cf., e.g., \citeNP{Leininger:1993}; \citeNP{Tabarrok/Spector:1999}; or \citeNP{Maskin/Sen:2016}).
We deem it worthwhile to evaluate collective choice methods also from an a~priori perspective and `on average'. % and to evaluate day-to-day affairs such as shareholder meetings.
For the simplest case with three options, we have considered the power implications of all conceivable weight distributions among three players and identified differences in how several prominent rules translate weights into voting power. % and derived related `rules of thumb' derived from . %For the five rules in our focus, e
Except for Borda rule, precise knowledge about the distribution of voting weights is typically not needed for gaging the sensitivity of outcomes to preferences of a large, middle, or small player respectively. % who should seek to maximize influence for the given distribution of weights.   
It is of course desirable to obtain results %on hidden biases of the rules 
also for bigger numbers of players and alternatives in the future.
% in order to learn more about %check the robustness of 
%that the adoption of a particular procedure for aggregating preferences on more than two alternatives may create. 
%should matter for the design of institutions and all committees whose decisions are not restricted to purely binary choices. 

Our analysis hopefully encourages the extension of voting power analysis to richer choice settings. 
There is a great variety of single-winner rules that could be added to the influence map in Figure~\ref{fig:Best_rule_3_3} (see, e.g., \citeNP{Aleskerov/Kurbanov:1999}; \citeNP[Ch.~7]{Nurmi:2006}; or \citeNP{Laslier:2012}). And nothing in principle would preclude similar analysis for multi-winner elections (see, e.g., \citeNP{Elkind/Faliszewski/Skowron/Slinko:2017}) or strategic voting equilibria, at least for restricted preference domains. 
It also seems worthwhile to investigate weight apportionment for two-tiered voting systems, such as US presidential elections, with more than two promising candidates. It is unknown, for instance, how the Penrose square root rule for the independent binary case (see, e.g, \citeNP[Ch.~3.4]{Felsenthal/Machover:1998}) or Shapley linear rule for affiliated spatial preferences %over intervals 
(cf.\ \citeNP{Kurz/Maaser/Napel:2018}) extend to two-tiered plurality decisions or plurality voting with a runoff. %with or without runoff, for instance. %We leave this for future research. 

\newpage

\setlength{\labelsep}{-0.2cm}

\phantomsection \label{References}
\addcontentsline{toc}{section}{References}

%%\bibliographystyle{chicago}
%%\bibliography{KMN-Literatur}

\begin{thebibliography}{}

\bibitem[\protect\citeauthoryear{Aleskerov, Kalyagin, and
  Pogorelskiy}{Aleskerov et~al.}{2008}]{Aleskerov/Kalyagin/Pogorelskiy:2008}
Aleskerov, F., V.~Kalyagin, and K.~Pogorelskiy (2008).
\newblock Actual voting power of the {IMF} members based on their
  political-economic integration.
\newblock {\em Mathematical and Computer Modelling\/}~{\em 48\/}(9--10),
  1554--1569.

\bibitem[\protect\citeauthoryear{Aleskerov and Kurbanov}{Aleskerov and
  Kurbanov}{1999}]{Aleskerov/Kurbanov:1999}
Aleskerov, F. and E.~Kurbanov (1999).
\newblock Degree of manipulability of social choice procedures.
\newblock In A.~Alkan, C.~D. Aliprantis, and N.~C. Yannelis (Eds.), {\em
  Current Trends in Economics}, pp.\  13--27. Berlin: Springer.

\bibitem[\protect\citeauthoryear{Amer, Carreras, and Mag{\~{a}}na}{Amer
  et~al.}{1998}]{Amer/Carreras/Magana:1998b}
Amer, R., F.~Carreras, and A.~Mag{\~{a}}na (1998).
\newblock Extension of values to games with multiple alternatives.
\newblock {\em Annals of Operations Research\/}~{\em 84\/}(0), 63--78.

\bibitem[\protect\citeauthoryear{Banzhaf}{Banzhaf}{1965}]{Banzhaf:1965}
Banzhaf, J.~F. (1965).
\newblock Weighted voting doesn't work: a mathematical analysis.
\newblock {\em Rutgers Law Review\/}~{\em 19\/}(2), 317--343.

\bibitem[\protect\citeauthoryear{Bolger}{Bolger}{1986}]{Bolger:1986}
Bolger, E.~M. (1986).
\newblock Power indices for multicandidate voting games.
\newblock {\em International Journal of Game Theory\/}~{\em 15\/}(3), 175--186.

\bibitem[\protect\citeauthoryear{Brams}{Brams}{1978}]{Brams:1978}
Brams, S.~J. (1978).
\newblock {\em The Presidential Election Game}.
\newblock New Haven, CT: Yale University Press.

\bibitem[\protect\citeauthoryear{Chua, Ueng, and Huang}{Chua
  et~al.}{2002}]{Chua/Ueng/Huang:2002}
Chua, V. C.~H., C.~H. Ueng, and H.~C. Huang (2002).
\newblock A method for evaluating the behavior of power indices in weighted
  plurality games.
\newblock {\em Social Choice and Welfare\/}~{\em 19\/}(3), 665--680.

\bibitem[\protect\citeauthoryear{Dubey and Shapley}{Dubey and
  Shapley}{1979}]{Dubey/Shapley:1979}
Dubey, P. and L.~Shapley (1979).
\newblock Mathematical properties of the {B}anzhaf power index.
\newblock {\em Mathematics of Operations Research\/}~{\em 4\/}(2), 99--131.

\bibitem[\protect\citeauthoryear{Elkind, Faliszewski, Skowron, and
  Slinko}{Elkind et~al.}{2017}]{Elkind/Faliszewski/Skowron/Slinko:2017}
Elkind, E., P.~Faliszewski, P.~Skowron, and A.~Slinko (2017).
\newblock Properties of multiwinner voting rules.
\newblock {\em Social Choice and Welfare\/}~{\em 48\/}(3), 599--632.

\bibitem[\protect\citeauthoryear{Felsenthal and Machover}{Felsenthal and
  Machover}{1997}]{Felsenthal/Machover:1997}
Felsenthal, D.~S. and M.~Machover (1997).
\newblock Ternary voting games.
\newblock {\em International Journal of Game Theory\/}~{\em 26\/}(3), 335--351.

\bibitem[\protect\citeauthoryear{Felsenthal and Machover}{Felsenthal and
  Machover}{1998}]{Felsenthal/Machover:1998}
Felsenthal, D.~S. and M.~Machover (1998).
\newblock {\em The Measurement of Voting Power}.
\newblock Cheltenham, UK: Edward Elgar.

\bibitem[\protect\citeauthoryear{Freixas}{Freixas}{2005a}]{Freixas:2005:Banzhaf}
Freixas, J. (2005a).
\newblock {Banzhaf} measures for games with several levels of approval in the
  input and output.
\newblock {\em Annals of Operations Research\/}~{\em 137\/}(1), 45--66.

\bibitem[\protect\citeauthoryear{Freixas}{Freixas}{2005b}]{Freixas:2005:Shapley}
Freixas, J. (2005b).
\newblock The {Shapley-Shubik} index for games with several levels of approval
  in the input and output.
\newblock {\em Decision Support Systems\/}~{\em 39\/}(2), 185--195.

\bibitem[\protect\citeauthoryear{Freixas and Zwicker}{Freixas and
  Zwicker}{2003}]{Freixas/Zwicker:2003}
Freixas, J. and W.~S. Zwicker (2003).
\newblock Weighted voting, abstention, and multiple levels of approval.
\newblock {\em Social Choice and Welfare\/}~{\em 21\/}(3), 399--431.

\bibitem[\protect\citeauthoryear{Freixas and Zwicker}{Freixas and
  Zwicker}{2009}]{Freixas/Zwicker:2009}
Freixas, J. and W.~S. Zwicker (2009).
\newblock Anonymous yes-no voting with abstention and multiple levels of
  approval.
\newblock {\em Games and Economic Behavior\/}~{\em 67\/}(2), 428--444.

\bibitem[\protect\citeauthoryear{Garrett and Tsebelis}{Garrett and
  Tsebelis}{1999}]{Garrett/Tsebelis:1999}
Garrett, G. and G.~Tsebelis (1999).
\newblock Why resist the temptation to apply power indices to the {European
  Union}?
\newblock {\em Journal of Theoretical Politics\/}~{\em 11\/}(3), 291--308.

\bibitem[\protect\citeauthoryear{Garrett and Tsebelis}{Garrett and
  Tsebelis}{2001}]{Garrett/Tsebelis:2001}
Garrett, G. and G.~Tsebelis (2001).
\newblock Even more reasons to resist the temptation of power indices in the
  {EU}.
\newblock {\em Journal of Theoretical Politics\/}~{\em 13\/}(1), 99--105.

\bibitem[\protect\citeauthoryear{Gibbard}{Gibbard}{1973}]{Gibbard:1973}
Gibbard, A. (1973).
\newblock Manipulation of voting schemes: a general result.
\newblock {\em Econometrica\/}~{\em 41\/}(4), 587--601.

\bibitem[\protect\citeauthoryear{Holler and Nurmi}{Holler and
  Nurmi}{2013}]{Holler/Nurmi:2013}
Holler, M.~J. and H.~Nurmi (Eds.) (2013).
\newblock {\em Power, Voting, and Voting Power: 30 Years After}.
\newblock Heidelberg: Springer.

\bibitem[\protect\citeauthoryear{Hsiao and Raghavan}{Hsiao and
  Raghavan}{1993}]{Hsiao/Raghavan:1993}
Hsiao, C.-R. and T.~E.~S. Raghavan (1993).
\newblock {Shapley} value for multichoice cooperative games, {I}.
\newblock {\em Games and Economic Behavior\/}~{\em 5\/}(2), 240--256.

\bibitem[\protect\citeauthoryear{Kelly}{Kelly}{1993}]{Kelly:1993}
Kelly, J.~S. (1993).
\newblock Almost all social choice rules are highly manipulable, but a few
  aren't.
\newblock {\em Social Choice and Welfare\/}~{\em 10\/}(2), 161--175.

\bibitem[\protect\citeauthoryear{Kurz, Maaser, and Napel}{Kurz
  et~al.}{2018}]{Kurz/Maaser/Napel:2018}
Kurz, S., N.~Maaser, and S.~Napel (2018).
\newblock Fair representation and a linear {S}hapley rule.
\newblock {\em Games and Economic Behavior\/}~{\em 108}, 152--161.

\bibitem[\protect\citeauthoryear{Kurz, Mayer, and Napel}{Kurz
  et~al.}{2019}]{Kurz/Mayer/Napel:2018}
Kurz, S., A.~Mayer, and S.~Napel (2019).
\newblock Weighted committee games.
\newblock {\em European Journal of Operational Research\/}, doi: 10.1016/j.ejor.2019.10.023.

\bibitem[\protect\citeauthoryear{Laruelle and Valenciano}{Laruelle and
  Valenciano}{2001}]{Laruelle/Valenciano:2001}
Laruelle, A. and F.~Valenciano (2001).
\newblock {Shapley-Shubik} and {Banzhaf} indices revisited.
\newblock {\em Mathematics of Operations Research\/}~{\em 26\/}(1), 89--104.

\bibitem[\protect\citeauthoryear{Laruelle and Valenciano}{Laruelle and
  Valenciano}{2008}]{Laruelle/Valenciano:2008}
Laruelle, A. and F.~Valenciano (2008).
\newblock {\em Voting and Collective Decision-Making}.
\newblock Cambridge: Cambridge University Press.

\bibitem[\protect\citeauthoryear{Laruelle and Valenciano}{Laruelle and
  Valenciano}{2012}]{Laruelle/Valenciano:2012}
Laruelle, A. and F.~Valenciano (2012).
\newblock Quaternary dichotomous voting rules.
\newblock {\em Social Choice and Welfare\/}~{\em 38\/}(3), 431--454.

\bibitem[\protect\citeauthoryear{Laslier}{Laslier}{2012}]{Laslier:2012}
Laslier, J.-F. (2012).
\newblock And the loser is \ldots plurality voting.
\newblock In D.~S. Felsenthal and M.~Machover (Eds.), {\em Electoral Systems:
  Paradoxes, Assumptions, and Procedures}, pp.\  327--351. Berlin: Springer.

\bibitem[\protect\citeauthoryear{Leech}{Leech}{2002}]{Leech:2002}
Leech, D. (2002).
\newblock Voting power in the governance of the {I}nternational {M}onetary
  {F}und.
\newblock {\em Annals of Operations Research\/}~{\em 109\/}(1--4), 375--397.

\bibitem[\protect\citeauthoryear{Leech}{Leech}{2003}]{Leech:2003}
Leech, D. (2003).
\newblock Computing power indices for large voting games.
\newblock {\em Management Science\/}~{\em 49\/}(6), 831--837.

\bibitem[\protect\citeauthoryear{Leech and Leech}{Leech and
  Leech}{2013}]{Leech/Leech:2013}
Leech, D. and R.~Leech (2013).
\newblock A new analysis of a priori voting power in the {IMF}: Recent quota
  reforms give little cause for celebration.
\newblock In M.~J. Holler and H.~Nurmi (Eds.), {\em Power, Voting, and Voting
  Power: 30 Years After}, pp.\  389--410. Heidelberg: Springer.

\bibitem[\protect\citeauthoryear{Leininger}{Leininger}{1993}]{Leininger:1993}
Leininger, W. (1993).
\newblock The fatal vote: {B}erlin versus {B}onn.
\newblock {\em Finanzarchiv\/}~{\em 50\/}(1), 1--20.

\bibitem[\protect\citeauthoryear{Mann and Shapley}{Mann and
  Shapley}{1962}]{Mann/Shapley:1962}
Mann, I. and L.~S. Shapley (1962).
\newblock Values of large games, {VI}: evaluating the {Electoral College}
  exactly.
\newblock Memorandum RM-3158-PR, The Rand Corporation.

\bibitem[\protect\citeauthoryear{Maskin and Sen}{Maskin and
  Sen}{2016}]{Maskin/Sen:2016}
Maskin, E. and A.~Sen (2016).
\newblock How majority rule might have stopped {Donald Trump}.
\newblock {\em New York Times\/}.
\newblock April 28, 2016.

\bibitem[\protect\citeauthoryear{Moulin}{Moulin}{1981}]{Moulin:1981}
Moulin, H. (1981).
\newblock The proportional veto principle.
\newblock {\em Review of Economic Studies\/}~{\em 48\/}(3), 407--416.

\bibitem[\protect\citeauthoryear{Napel}{Napel}{2019}]{Napel:2018}
Napel, S. (2019).
\newblock Voting power.
\newblock In R.~Congleton, B.~Grofman, and S.~Voigt (Eds.), {\em Oxford
  Handbook of Public Choice}, Volume~1, Chapter~6, pp.\  103--126. Oxford:
  Oxford University Press.

\bibitem[\protect\citeauthoryear{Napel and Widgr\'{e}n}{Napel and
  Widgr\'{e}n}{2004}]{Napel/Widgren:2004}
Napel, S. and M.~Widgr\'{e}n (2004).
\newblock Power measurement as sensitivity analysis: a unified approach.
\newblock {\em Journal of Theoretical Politics\/}~{\em 16\/}(4), 517--538.

\bibitem[\protect\citeauthoryear{Nitzan}{Nitzan}{1985}]{Nitzan:1985}
Nitzan, S. (1985).
\newblock The vulnerability of point-voting schemes to preference variation and
  strategic manipulation.
\newblock {\em Public Choice\/}~{\em 47\/}(2), 349--370.

\bibitem[\protect\citeauthoryear{Nurmi}{Nurmi}{2006}]{Nurmi:2006}
Nurmi, H. (2006).
\newblock {\em Models of Political Economy}.
\newblock London: Routledge.

\bibitem[\protect\citeauthoryear{Owen}{Owen}{1975}]{Owen:1975:presidential}
Owen, G. (1975).
\newblock Evaluation of a presidential election game.
\newblock {\em American Political Science Review\/}~{\em 69\/}(3), 947--953.

\bibitem[\protect\citeauthoryear{Parker}{Parker}{2012}]{Parker:2012}
Parker, C. (2012).
\newblock The influence relation for ternary voting games.
\newblock {\em Games and Economic Behavior\/}~{\em 75\/}(2), 867--881.

\bibitem[\protect\citeauthoryear{Peleg}{Peleg}{1984}]{Peleg:1984}
Peleg, B. (1984).
\newblock {\em Game Theoretic Analysis of Voting in Committees}.
\newblock Cambridge: Cambridge University Press.

\bibitem[\protect\citeauthoryear{Penrose}{Penrose}{1946}]{Penrose:1946}
Penrose, L.~S. (1946).
\newblock The elementary statistics of majority voting.
\newblock {\em Journal of the Royal Statistical Society\/}~{\em 109\/}(1),
  53--57.

\bibitem[\protect\citeauthoryear{Regenwetter, Grofman, Marley, and
  Tsetlin}{Regenwetter et~al.}{2012}]{Regenwetter/Grofman/Marley/Tsetlin:2012}
Regenwetter, M., B.~Grofman, A.~A.~J. Marley, and I.~M. Tsetlin (2012).
\newblock {\em Behavioral Social Choice\/} (2nd edition ed.).
\newblock New York, NY: Cambridge University Press.

\bibitem[\protect\citeauthoryear{Riker and Shapley}{Riker and
  Shapley}{1968}]{Riker/Shapley:1968}
Riker, W.~H. and L.~S. Shapley (1968).
\newblock Weighted voting: a mathematical analysis for instrumental judgements.
\newblock In J.~R. Pennock and J.~W. Chapman (Eds.), {\em Representation:
  Nomos~X}, Yearbook of the American Society for Political and Legal
  Philosophy, pp.\  199--216. New York, NY: Atherton Press.

\bibitem[\protect\citeauthoryear{Satterthwaite}{Satterthwaite}{1975}]{Satterthwa
ite:1975}
Satterthwaite, M.~A. (1975).
\newblock Strategy-proofness and {A}rrow's conditions: existence and
  correspondence theorems for voting procedures and social welfare function.
\newblock {\em Journal of Economic Theory\/}~{\em 10\/}(2), 187--217.

\bibitem[\protect\citeauthoryear{Schulze}{Schulze}{2011}]{Schulze:2011}
Schulze, M. (2011).
\newblock A new monotonic, clone-independent, reversal symmetric, and
  condorcet-consistent single-winner election method.
\newblock {\em Social Choice and Welfare\/}~{\em 36\/}(2), 267--303.

\bibitem[\protect\citeauthoryear{Shapley and Shubik}{Shapley and
  Shubik}{1954}]{Shapley/Shubik:1954}
Shapley, L.~S. and M.~Shubik (1954).
\newblock A method for evaluating the distribution of power in a committee
  system.
\newblock {\em American Political Science Review\/}~{\em 48\/}(3), 787--792.

\bibitem[\protect\citeauthoryear{Smith}{Smith}{1999}]{Smith:1999}
Smith, D.~A. (1999).
\newblock Manipulability measures of common social choice functions.
\newblock {\em Social Choice and Welfare\/}~{\em 16\/}(4), 639--661.

\bibitem[\protect\citeauthoryear{Tabarrok and Spector}{Tabarrok and
  Spector}{1999}]{Tabarrok/Spector:1999}
Tabarrok, A. and L.~Spector (1999).
\newblock Would the {B}orda {C}ount have avoided the civil war?
\newblock {\em Journal of Theoretical Politics\/}~{\em 11\/}(2), 261--288.

\bibitem[\protect\citeauthoryear{Tchantcho, Lambo, Pongou, and
  Engoulou}{Tchantcho et~al.}{2008}]{Tchantcho/DiffoLambo/Pongou/Engoulou:2008}
Tchantcho, B., L.~D. Lambo, R.~Pongou, and B.~M. Engoulou (2008).
\newblock Voters' power in voting games with abstention: influence relation and
  ordinal equivalence of power theories.
\newblock {\em Games and Economic Behavior\/}~{\em 64\/}(1), 335--350.

\bibitem[\protect\citeauthoryear{Von~Neumann and Morgenstern}{Von~Neumann and
  Morgenstern}{1953}]{vonNeumann/Morgenstern:1953}
Von~Neumann, J. and O.~Morgenstern (1953).
\newblock {\em Theory of Games and Economic Behavior\/} (3rd ed.).
\newblock Princeton, NJ: Princeton University Press.
\newblock %(1st edition, 1944).

\end{thebibliography}

\newpage

\section*{Appendix: Comparisons of voting rules for \emph{m}$\,\mathbf{=3}$}
\phantomsection 
\addcontentsline{toc}{section}{Appendix}

\vspace{-1cm}
\renewcommand{\thefigure}{A-\arabic{figure}}
\setcounter{figure}{0}

% HINWEIS: max width damit 2 Abbildungen auf eine Seite passen: 0.69
\begin{figure}[h!]
\centering
  \includegraphics[width=0.69\textwidth]{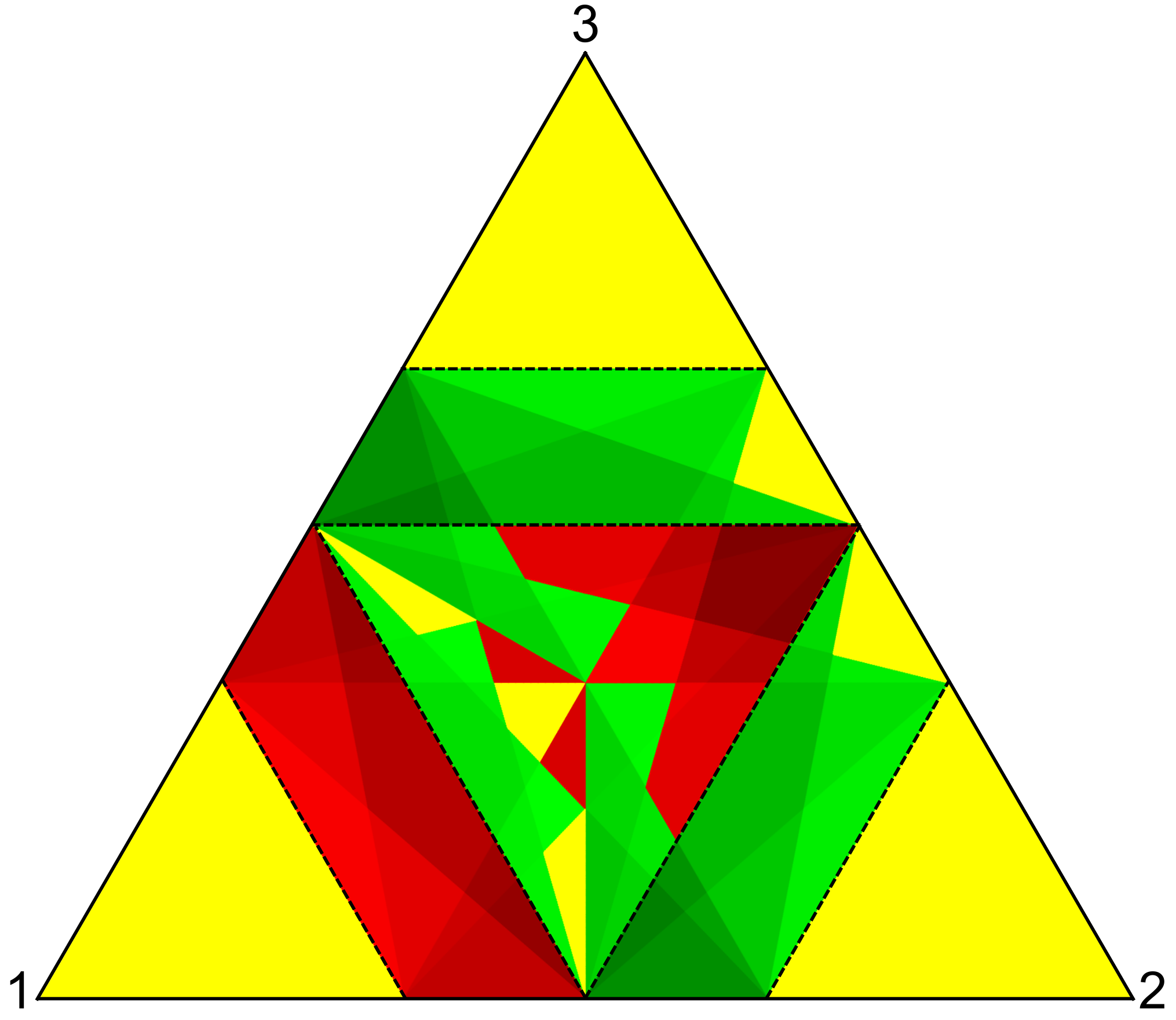}
      \caption{\small Borda vs. plurality (repeated from p.~\pageref{fig:Borda_Plurality_3_3_example}) 
\label{fig:Borda_Plurality_3_3}}
\end{figure}

%\vspace*{\floatsep}% https://tex.stackexchange.com/q/26521/5764
\vspace{-.7cm}

\begin{figure}[h!]
\centering
  \includegraphics[width=0.69\textwidth]{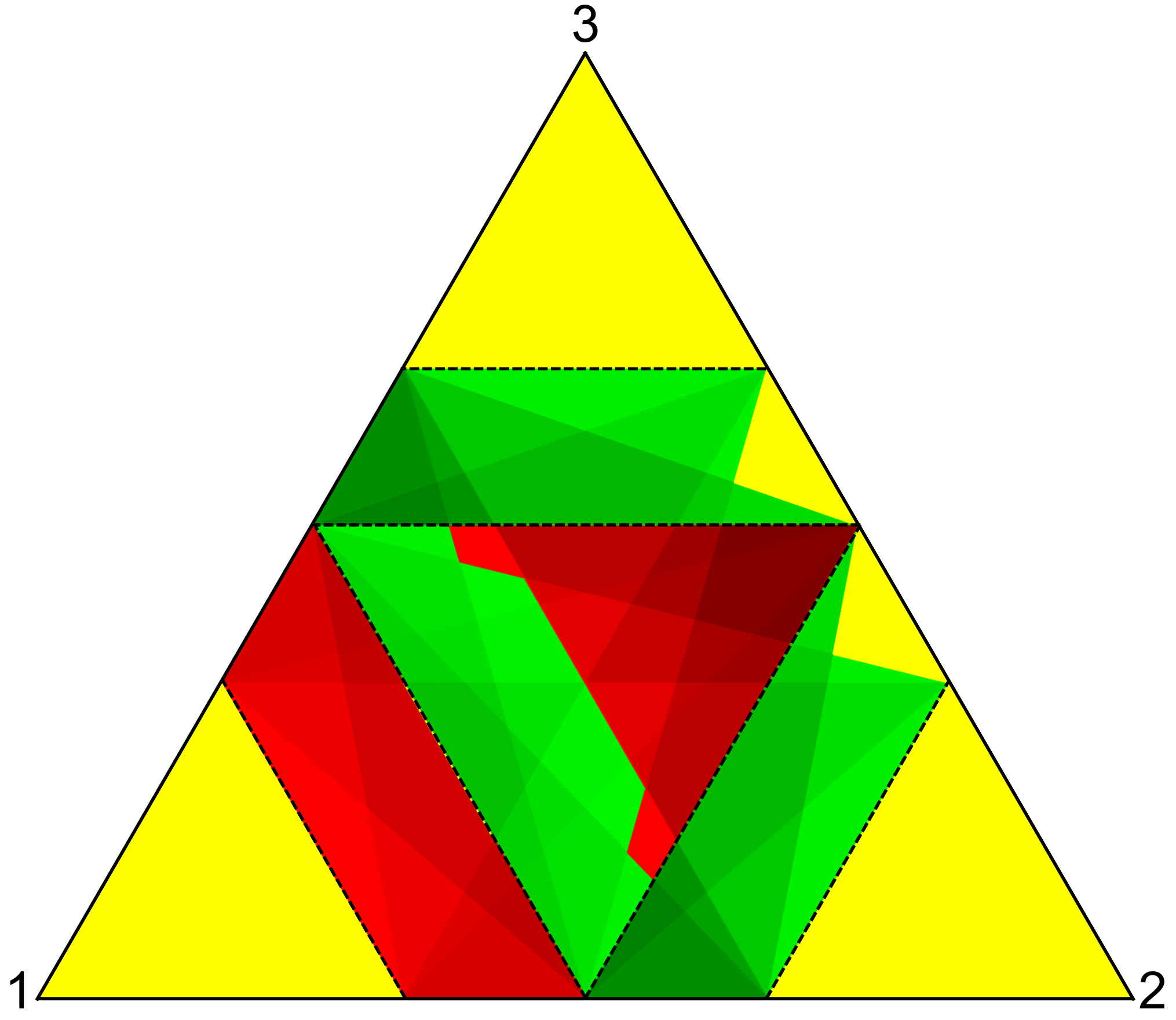}
      \caption{\small Borda vs. Copeland 
\label{fig:Borda_Copeland_3_3}}
\end{figure}

\newpage

\begin{figure}[h!]
\centering
  \includegraphics[width=0.69\textwidth]{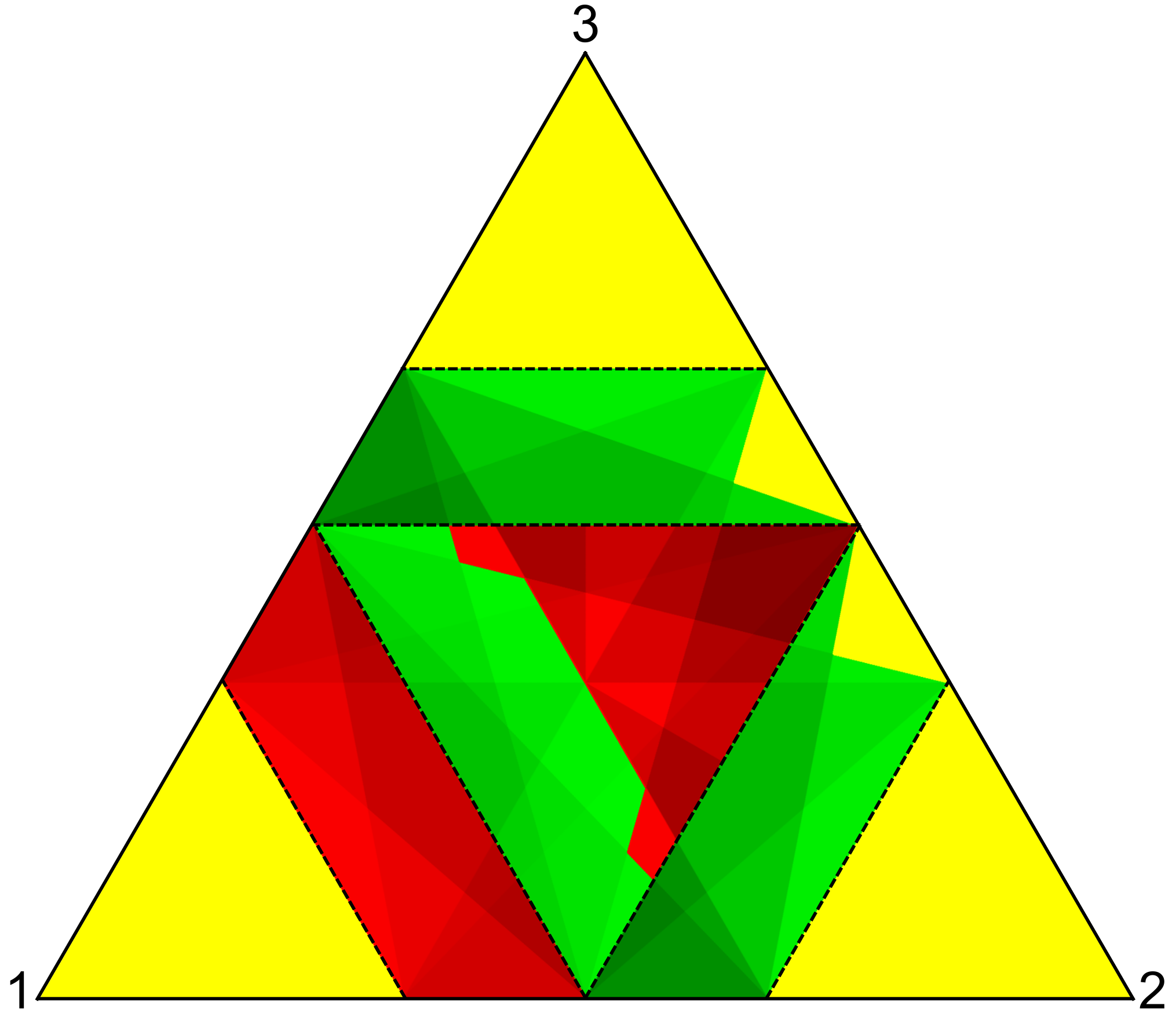}
      \caption{\small Borda vs. plurality runoff 
\label{fig:Borda_PluralityRunoff_3_3}}
\end{figure}

\vspace*{\floatsep}% https://tex.stackexchange.com/q/26521/5764

\begin{figure}[h!]
\centering
  \includegraphics[width=0.69\textwidth]{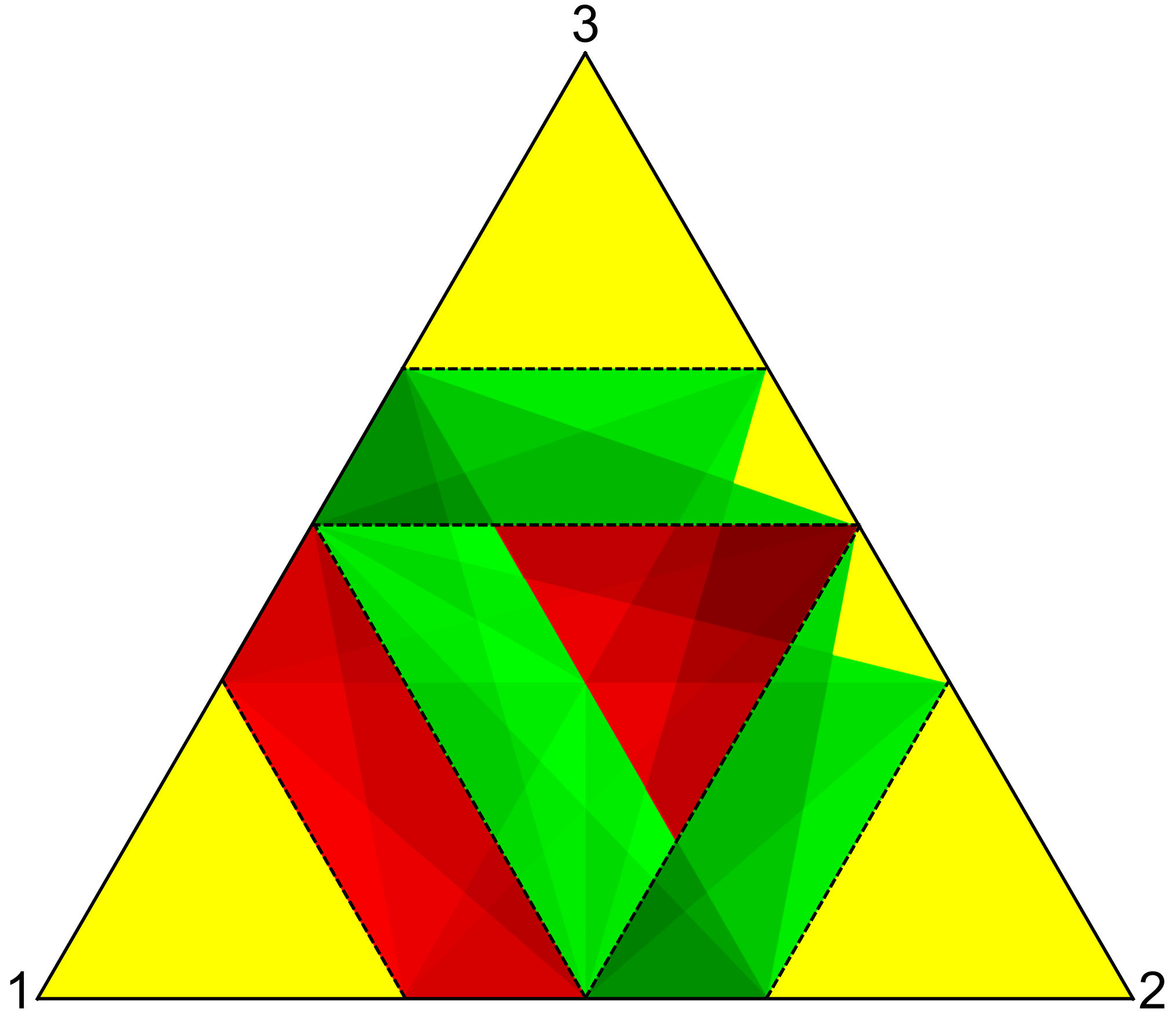}
      \caption{\small Borda vs. Schulze
\label{fig:Borda_Schulze_3_3}}
\end{figure}

\newpage 

%\begin{figure}[htbp]
%\centering
%  \includegraphics[width=0.69\textwidth]{Plurality_Copeland_3_3.pdf}
%      \caption{\small Plurality vs. Copeland for %$n=3$ and 
%$m=3$\label{fig:Plurality_PluralityRunoffCopeland/Schulze_3_3}}

\begin{figure}[h!]
\centering
  \includegraphics[width=0.69\textwidth]{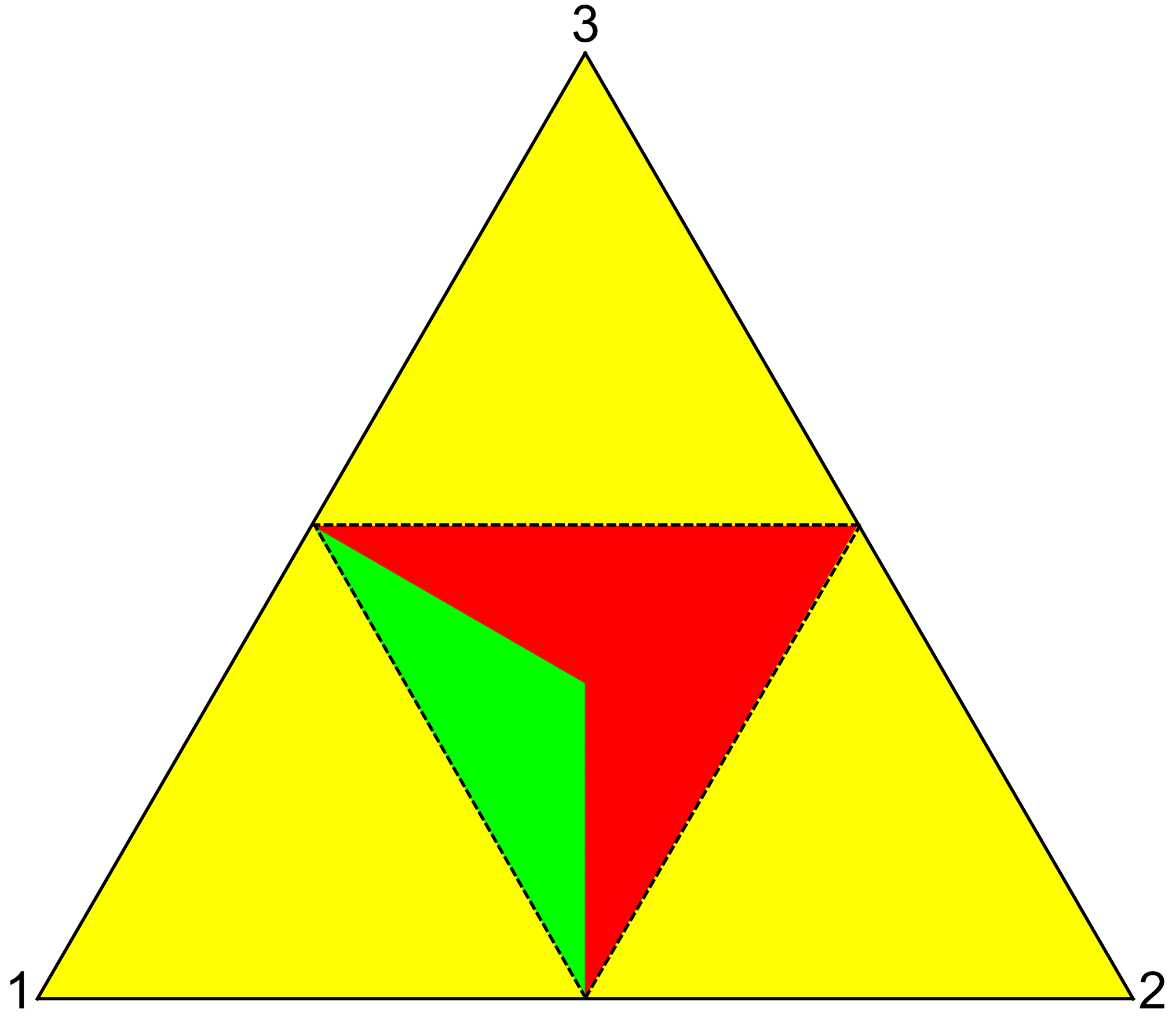}
      \caption{\small Plurality vs. plurality runoff, Copeland and Schulze
\label{fig:Plurality_PluralityRunoffCopelandSchulze_3_3}}
\end{figure}

\vspace*{\floatsep}% https://tex.stackexchange.com/q/26521/5764

%\centering
%  \includegraphics[width=0.69\textwidth]{Plurality_PluralityRunoff_3_3.pdf}
%      \caption{\small Plurality vs. Plurality runoff 
%\label{fig:Plurality_PluralityRunoff_3_3}}
%\end{figure}

%\newpage

%\begin{figure}[htbp]
%\centering
%  \includegraphics[width=0.69\textwidth]{Plurality_Schulze_3_3.pdf}
%      \caption{\small Plurality vs. Schulze 
%\label{fig:Plurality_Schulze_3_3}}

%\vspace*{\floatsep}% https://tex.stackexchange.com/q/26521/5764

\begin{figure}[h!]
\centering
  \includegraphics[width=0.69\textwidth]{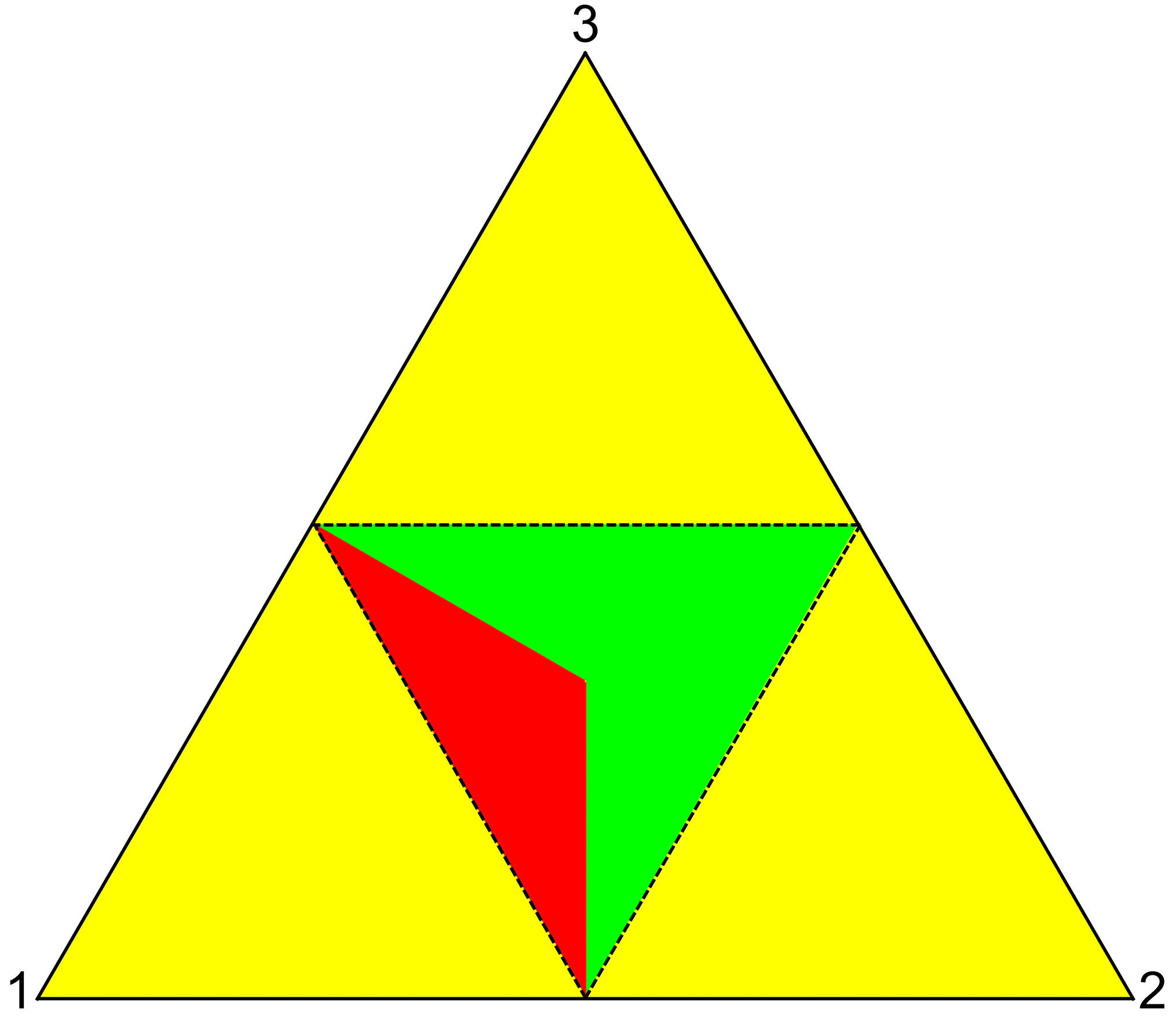}
      \caption{\small Copeland vs. Schulze
\label{fig:Copeland_Schulze_3_3}}
\end{figure}

\newpage

\begin{figure}[h!]
\centering
  \includegraphics[width=0.69\textwidth]{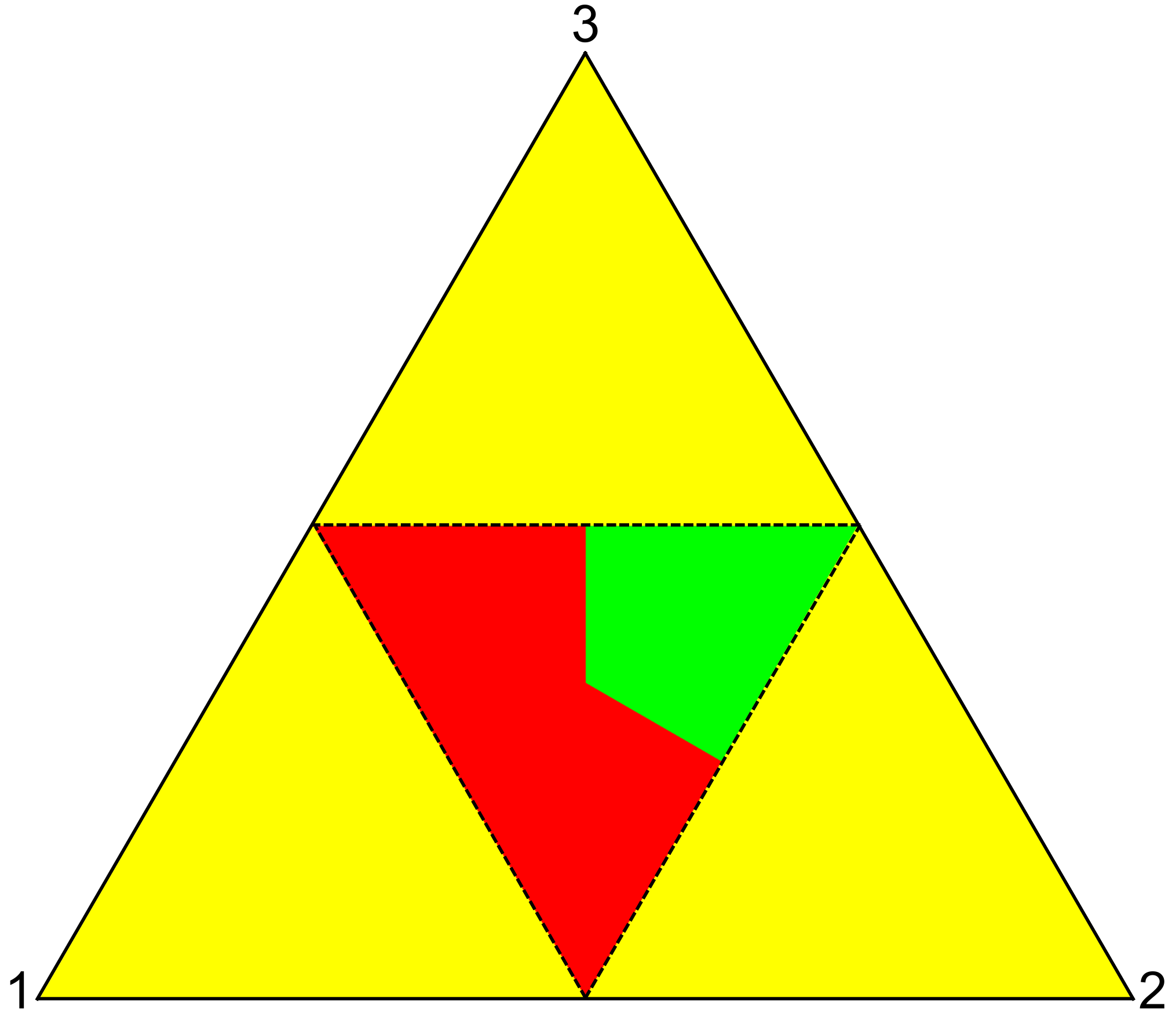}
      \caption{\small Copeland vs. plurality runoff
\label{fig:Copeland_PluralityRunoff_3_3}}
\end{figure}

\vspace*{\floatsep}% https://tex.stackexchange.com/q/26521/5764

\begin{figure}[h!]
\centering
  \includegraphics[width=0.69\textwidth]{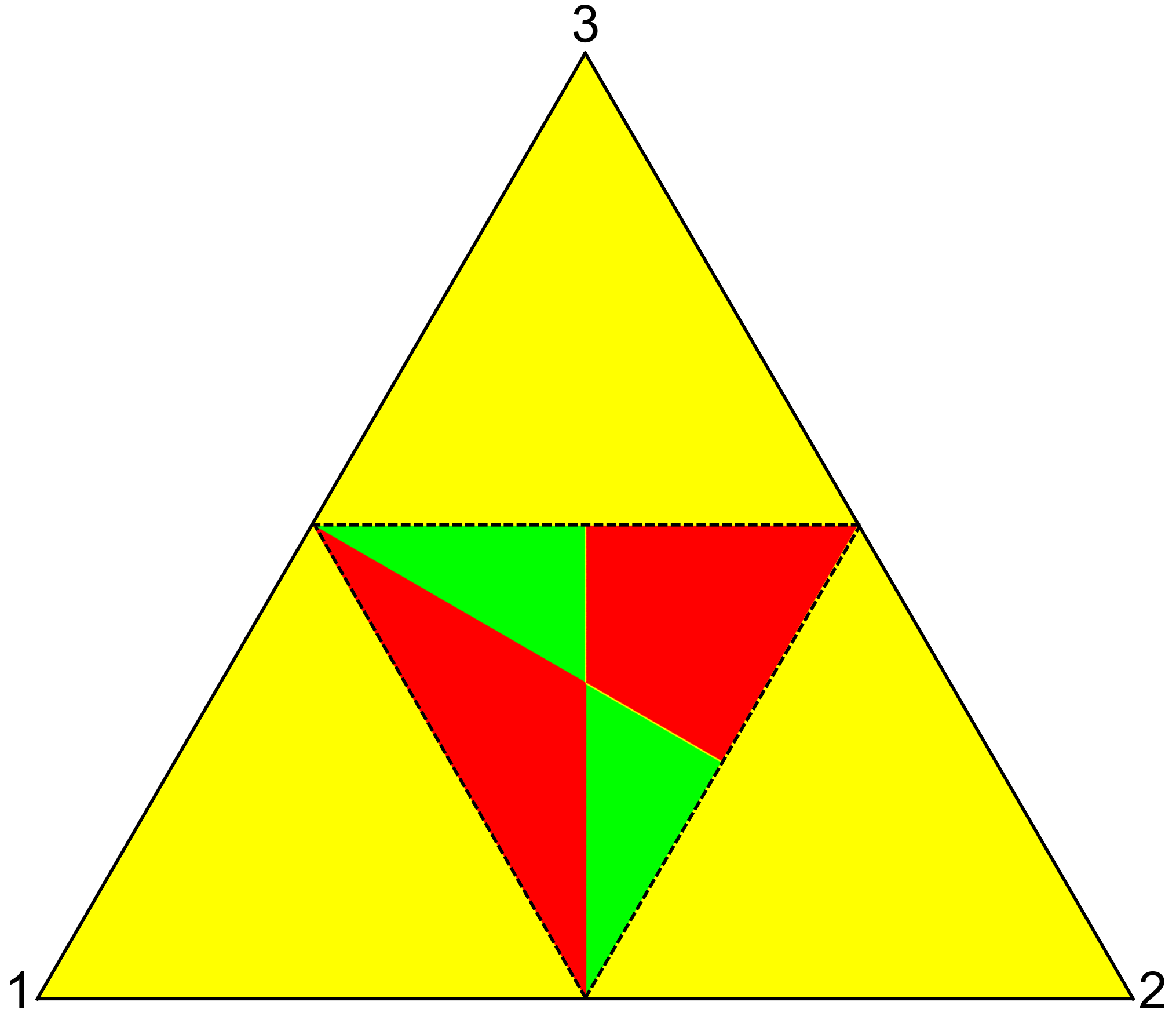}
      \caption{\small Plurality runoff vs. Schulze
\label{fig:PluralityRunoff_Schulze_3_3}}
\end{figure}

\end{document}